\documentclass{aa}  

\usepackage{graphicx}
\usepackage{txfonts}
\usepackage{subcaption}         
\usepackage{lscape}             
\usepackage{placeins}           
                                
\usepackage[colorlinks=true, citecolor=blue, linkcolor=blue, urlcolor=blue]{hyperref} 

\def\msun{\ifmmode M_{\odot} \else M$_{\odot}$\fi}
\def\msunyr{\ifmmode M_{\odot} {\rm yr}^{-1} \else M$_{\odot}$ yr$^{-1}$\fi}
\def\zsun{\ifmmode Z_{\odot} \else Z$_{\odot}$\fi}
\def\lsun{\ifmmode L_{\odot} \else L$_{\odot}$\fi}

\newcommand{\mstar}{\ifmmode M_\star \else $M_\star$\fi}
\newcommand{\muv}{\ifmmode M_{\rm UV} \else $M_{\rm UV}$\fi}
\newcommand{\auv}{\ifmmode A_{\rm UV} \else $A_{\rm UV}$\fi}
\newcommand{\luv}{\ifmmode L_{\rm UV} \else $L_{\rm UV}$\fi}
\newcommand{\lir}{\ifmmode L_{\rm IR} \else $L_{\rm IR}$\fi}
\newcommand{\lbol}{\ifmmode L_{\rm bol} \else $L_{\rm bol}$\fi}
\newcommand{\liruv}{\ifmmode L_{\rm IR+UV} \else $L_{\rm IR+UV}$\fi}
\newcommand{\liroveruv}{\ifmmode L_{\rm IR}/L_{\rm UV} \else $L_{\rm IR}/L_{\rm UV}$\fi}
\newcommand{\nlyc}{\ifmmode N_{\rm Lyc} \else $N_{\rm Lyc} $\fi}
\newcommand{\rholyc}{\ifmmode \rho_{\rm Lyc} \else $\rho_{\rm Lyc} $\fi}
\newcommand{\chion}{\ifmmode \xi_{\rm ion} \else $\xi_{\rm ion}$\fi}
\newcommand{\chioncorr}{\ifmmode \xi_{\rm ion}^0 \else $\xi_{\rm ion}^0$\fi}

\newcommand{\Heiiopt}{He~{\sc ii} $\lambda$4686}

\newcommand{\Heii}{He~{\sc ii}}

\newcommand{\Nii}{[N~{\sc ii}]}

\newcommand{\Oii}{[O~{\sc ii}]}
\newcommand{\Oiii}{[O~{\sc iii}]}
\newcommand{\Oiv}{[O~{\sc iv}]}
\newcommand{\Neii}{[Ne~{\sc ii}]}
\newcommand{\Neiii}{[Ne~{\sc iii}]}

\newcommand{\Arii}{[Ar~{\sc ii}]}
\newcommand{\Nev}{[Ne~{\sc v}]}

\newcommand{\Siii}{[S~{\sc iii}]}
\newcommand{\Siv}{[S~{\sc iv}]}

\newcommand{\Oi}{[O~{\sc i}]}
\newcommand{\Ha}{H$\alpha$}
\newcommand{\Hb}{\ifmmode {{\rm H}\beta} \else H$\beta$\fi}

\newcommand{\sbs}{SBS 0335-052 E}

\begin{document}

   \title{Investigating the ionization mechanisms powering the extreme emission lines in metal-poor galaxies
   }


%

\author{C. Daoutis\inst{1}\thanks{E-mail: Charalampos.Daoutis@unige.ch}, 
V. Lebouteiller\inst{2},
D. Schaerer\inst{1,3},
I. Morel\inst{1}, 
O. Bait\inst{4,5},
S. Satyapal\inst{6}, 
S. Doan\inst{6},
B. Trahin\inst{2},
A. Gurpide\inst{7}, 
Y. Izotov\inst{8},
N. Guseva\inst{8},
L. Ramambason\inst{9},
A. Verhamme\inst{1},
and SpeXion Team\inst{1,2,3}}
  \institute{Department of Astronomy, University of Geneva, Chemin Pegasi 51, 1290 Versoix, Switzerland
\and Université Paris-Saclay, Université Paris Cité, CEA, CNRS, AIM, 91191, Gif-sur-Yvette, France
\and CNRS, IRAP, 14 Avenue E. Belin, 31400 Toulouse, France
\and National Radio Astronomy Observatory, 520 Edgemont Road, Charlottesville, VA 22903, USA 
\and The NSF-Simons AI Institute for Cosmic Origins, USA, 201 E. 24th Street, POB 4.102, Austin, Texas 78712-1229 
\and Department of Physics and Astronomy, George Mason University, MS3F3, 4400 University Drive, Fairfax, VA 22030, USA
\and Anton Pannekoek Institute for Astronomy, University of Amsterdam, Science Park 904, 1098 XH Amsterdam, The Netherlands
\and Bogolyubov Institute for Theoretical Physics, National Academy of Sciences of Ukraine, 14-b Metrolohichna str., Kyiv, 03143, Ukraine
\and Institut fur Theoretische Astrophysik, Zentrum für Astronomie, Universität Heidelberg, Albert-Ueberle-Str. 2, 69120 Heidelberg, Germany
}

\authorrunning{Daoutis et al.}
\titlerunning{Ionization mechanisms in metal-poor galaxies}

   \date{Received date / Accepted date}

 
  \abstract
   {Extreme emission metal-poor galaxies (EEMPGs) serve as excellent local laboratories for understanding the extreme physical conditions and hard ionizing radiation fields prevalent in the early Universe. However, the exact source of high-ionization emission lines in them, whether from intermediate-mass black holes (IMBHs), ultra-luminous X-ray sources (ULXs), or radiative shocks, remains a subject of debate.}
   {We characterize the fundamental properties and the ionizing radiation field on metal-poor extreme emission-line galaxies. Specifically, we aim to identify the dominant sources of the highly energetic ionizing photons that drive these extreme emission lines.} 
   {We selected a sample of 11 EEMPGs, which are among the most metal-poor galaxies with JWST/MIRI observations known so far (12+log(O/H) < 8.0, or Z/Z$_\odot$ < 20\%). By using optical and infrared diagnostic diagrams for ions of both high and low ionization potential, we try to constrain the ionization source dominating their observed spectral features. Furthermore, using state-of-the-art photoionization models, we attempt to distinguish between radiative shocks, IMBHs, and ULXs as the primary drivers of the high-ionization emission.}
   {We detect high-ionization lines, notably \Heiiopt\ in all galaxies and \Nev 14.3\,$\mu$m in 42\% of a broader sample of local EEMPGs (in our sample and including objects from the literature). The multi-wavelength diagnostic diagrams reveal that our EEMPGs are located in the areas that normally are occupied by active galactic nuclei (AGNs) or between the star-formation and AGN sequences. When evaluating the energetics required to produce these lines, we find that radiative shocks and ULXs are not sufficient to explain the observed line fluxes and ratios. Instead, our analysis of the ionizing spectrum suggests that a combination of sources with star-formation and a small contribution from an IMBH provide enough high-energy photon budget to explain the emission lines of ions with high-ionization potentials (> 54 eV).}
   {Our EEMPG sample spans a range of properties, making it ideal for studying the first galaxies. Their flux ratios of low- and high-ionization lines suggest that their likely ionization source is stellar radiation with small a contribution of an IMBH (4--8\%), while pure shocks and ULXs fall short of explaining the observed emission line ratios and ionizing photon budget.}


   \maketitle
\nolinenumbers

\section{Introduction}

The unprecedented capabilities of JWST have revolutionized our understanding of galaxy evolution. Now, for the first time, it is possible to observe very faint galaxies across a wide range of distances. This has led to the discovery of many faint active galactic nuclei (AGN) in the Early Universe \citep[e.g.,][]{2023ApJ...959...39H,2023Natur.616..266L,2024A&A...691A.145M,2024Natur.636..594J,2024MNRAS.531..355U,2025ApJ...986..126K}. However, their role in the re-ionization of the Universe is still unclear. Over the past years, many studies have argued that low-mass, low-metallicity dwarf starburst galaxies played a crucial role in the re-ionization of the Universe \citep[e.g.,][]{2014MNRAS.442.2560W,2015ApJ...802L..19R,2015MNRAS.450.1846S}, a result that is further supported by recent observations \citep[e.g.,][]{2024Natur.626..975A,2024MNRAS.527.6139S}.

Understanding the conditions of galaxy evolution in the Early Universe relies on our study of dwarf, extremely metal-deficient (XMD; metallicity $\lesssim$ 10\% Z$_{\odot}$) galaxies \citep[stellar mass $\lesssim$ 10$^9$ $M_{\odot}$, see][]{2018PhR...780....1D,2024MNRAS.527.6139S,2000A&ARv..10....1K}, which are the most common galaxy type across all redshifts \citep[e.g.,][]{1991ApJ...379...52W,1996ApJ...458..100B,2017ARA&A..55..343B,2017A&A...605A..70D,2021ApJ...922...29S}. Furthermore, recent JWST studies have revealed that these galaxies exhibit several characteristics commonly observed in high-redshift galaxies, such as low metal abundances and elevated star formation rates \citep[e.g.,][]{2022A&A...665L...4S,2023MNRAS.525.2087B}, making them valuable analogs for studying galaxy evolution in the early Universe. Intense star-forming galaxies at low redshift offer ideal laboratories for studying interstellar medium (ISM) properties in quasi-pristine conditions, akin to the epoch of re-ionization \citep{2022Galax..10...11H}, particularly compact galaxies with hard ionization fields \citep{2022MNRAS.516L..81T,2024MNRAS.527.3486I}. Thus, these counterparts in the local Universe are invaluable for exploring feedback mechanisms and star formation under conditions similar to those of high-redshift galaxies. 

The presence of high ionization potential (> 54 eV) spectral lines in metal-poor galaxies indicates the existence of hard ionization fields, typically beyond what stellar populations can produce \citep{2026arXiv260904315B}. \Heii\ $\lambda$4686\ line is thought to be originating from Wolf-Rayet (WR) stars \citep{1996ApJ...467L..17S,1998ApJ...497..618S,1999A&AS..136...35S} and X-ray binaries (XRBs) \citep{1991ApJ...373..458G,2019A&A...622L..10S}, as well as from accreting black holes \citep{2012MNRAS.421.1043S}. However, higher ionization emission lines (e.g., \Nev14\,$\mu$m, with an ionization potential, I.P., of 97 eV) require even harder ionization fields, i.e., AGN \citep[e.g., see][]{1992ApJ...399..504S}. 

Fast radiative shocks propagating through the interstellar medium have also been proposed as the origin of these high ionization lines. Supernova explosions and winds accelerate gas to supersonic velocities, triggering shocks that collide with interstellar material, heating and ionizing the gas, causing photon emission and kinetic energy transference \citep{1995ApJ...455..468D, 1996ApJS..102..161D}, leading to the observed high ionization lines in galaxy spectra \citep{1996ApJS..102..161D,2004A&A...415L..27I,2013A&A...558A..57I}. Therefore, shock waves have also been used to explain high ionization emission lines \Nev\ and \Heii\ observed in local galaxies \citep{2005ApJS..161..240T,2012MNRAS.427.1229I,2021MNRAS.508.2556I}.

Optical activity diagnostics \citep{1981PASP...93....5B,2001ApJ...556..121K,2003MNRAS.346.1055K} are often employed to identify the dominant ionization source inside a galaxy. However, these are often prone to reddening effects as well as degeneracies in extremely low metallicity environments. Extension of these diagnostics to other lines, e.g., \Oi $\lambda$6300, can be a good indication of excitation from shocks. Mid-infrared diagnostics, mostly introduced during the \textit{Spitzer} Space Telescope (SST) era, can be use to distinguish between different galactic emission sources, such as star formation or AGN. Key diagnostics include the presence and intensity of high-ionization (and coronal lines) lines, such as \Nev\ 14.3\,$\mu$m, and serve as robust indicators of AGN activity, as these lines require hard ionizing photons not produced by stellar populations \citep{1992ApJ...399..504S,2009ApJ...693.1821D,2006A&A...446..877M,2010ApJ...716.1151W}.

Despite their importance, a comprehensive study of extremely metal-poor dwarf galaxies with hard ionization fields has not yet been performed. In this work, we present a sample of local, extremely metal-poor, compact star-forming galaxies with hard ionization fields, five of which are the most metal-poor known so far (12 + log(O/H) = 6.98--7.23). This is the first sample with complete mid-IR (5--25\,$\mu$m) JWST/MIRI spectra. In this work, we present physical properties such as metallicities,
stellar masses, and star-formation rates, obtained from optical observations, and we compare them against other literature samples of normal and compact extreme star-forming galaxies. Furthermore, we use the JWST observations to place these galaxies on mid-IR activity diagnostics in an attempt to explain the potential source of both their moderate and high ionization lines.

This paper is organized as follows. Section \ref{Sec_sample} outlines the physical properties of our galaxy sample, while Section \ref{Sec_JWST_obs} describes the JWST observations and data reduction. Section \ref{Sec_MIR_diagnostics} examines the location of our galaxies within standard mid-IR activity diagnostics, comparing their positions against representative literature samples of various activity types. In Section \ref{Sec_Origin_of_high_IP_lines}, we discuss possible excitation mechanisms that explain simultaneously the low- and high-ionization lines. Finally, Section \ref{Sec_conclusions} summarizes our conclusions.

\begin{figure}[htb]
  \resizebox{\hsize}{!}{\includegraphics{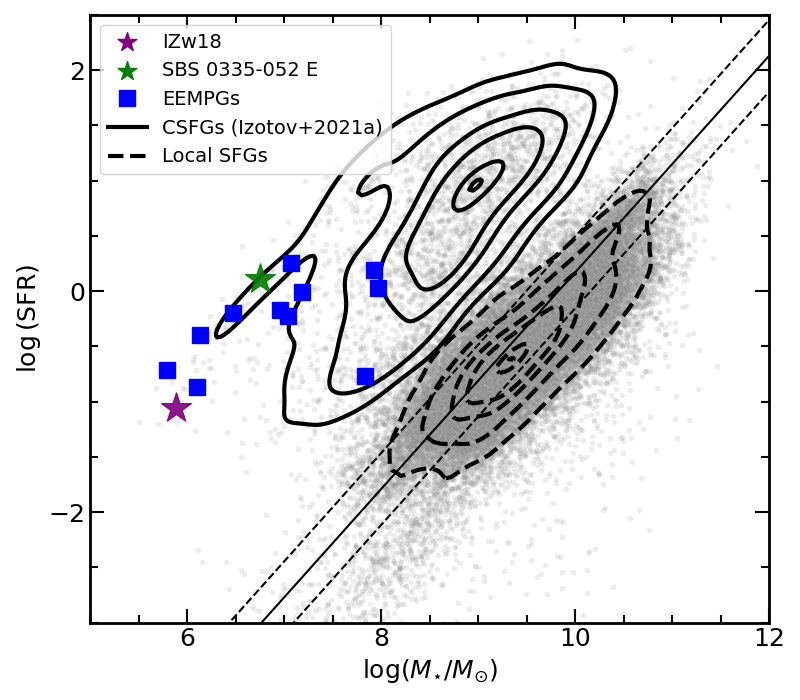}}
  \caption{Star formation rate (SFR) as a function of stellar mass ($M_*$). The dashed black contours outline the positions of typical local star-forming galaxies (main sequence, MS). Solid black contours indicate the distribution of the compact star-forming galaxy (CSFGs) sample \citep{2021A&A...646A.138I}, which show higher SFRs (and also have higher specific SFRs, i.e., SFR/$M_*$). Our EEMPGs (blue squares) are plotted alongside the well-known metal-deficient dwarfs I Zw 18 \citep[purple star;][]{2012MNRAS.427..906H} and SBS 0335-052 E \citep[green star;][]{2012MNRAS.427..906H}. The EEMPGs predominantly occupy the region well above the main sequence (MS) and the tail end of CSFGs, characterized by low $M_*$, which indicates high sSFRs typical of active star-forming metal-poor galaxies. The black solid line corresponds to the main sequence of star-forming galaxies and the black dashed lines on either side of it are the 1$\sigma$ to that fit \citep{2017MNRAS.466.1192M}. We also show a representative sample of local star-forming galaxies \citep[HECATE][]{2021MNRAS.506.1896K,2026MNRAS.548ag522K}.}
  \label{MS}
\end{figure}

\begin{figure}[htb]
  \resizebox{\hsize}{!}{\includegraphics{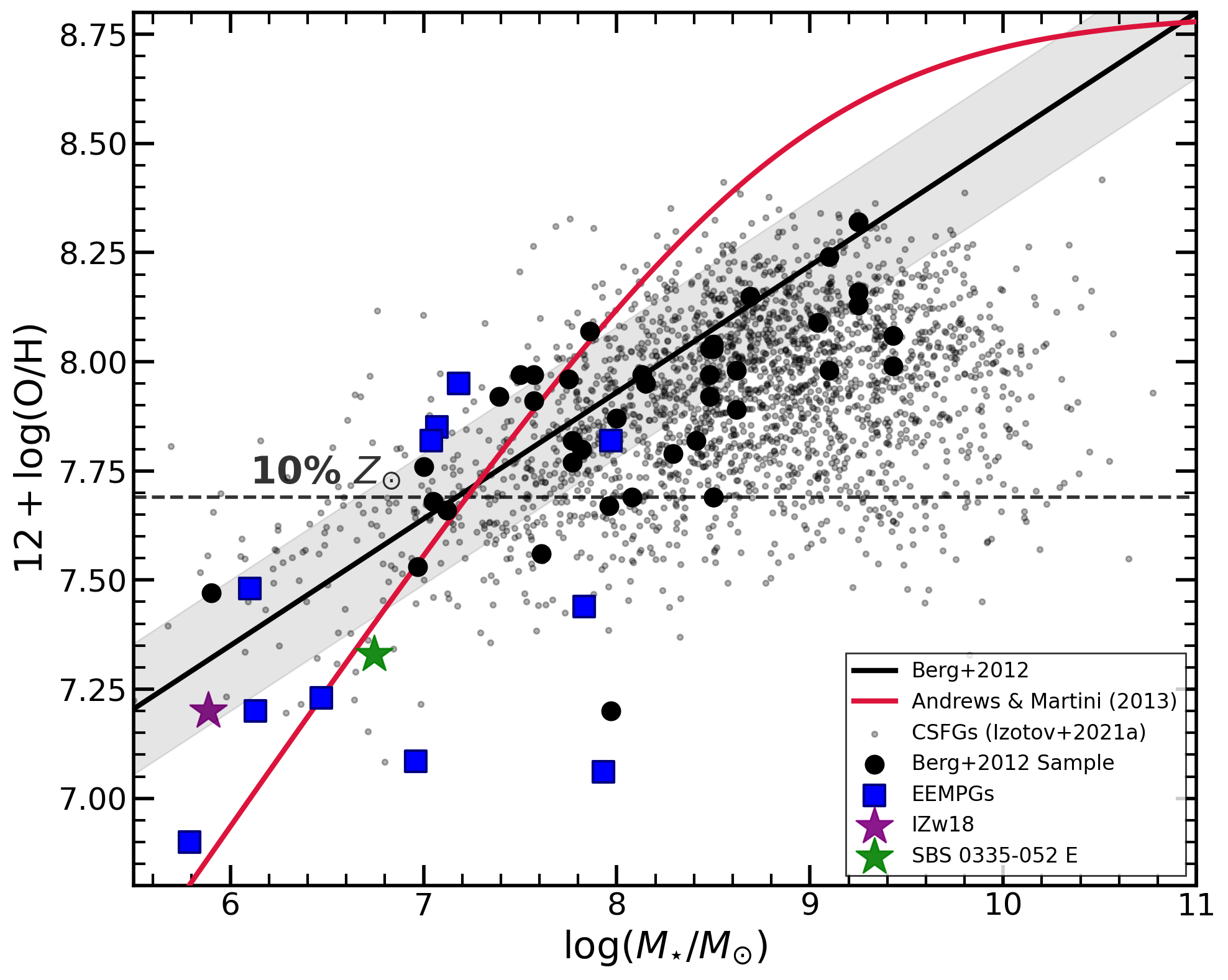}}
  \caption{Gas-phase oxygen abundance, expressed as $12 + \log(\text{O/H})$, is plotted as a function of stellar mass, $\log(M_* / M_{\odot})$ (mass-metallicity relation, MZR) for our EEMPGs (blue squares). Comparison samples include the CSFGs from \citet{2021A&A...646A.138I} (small gray circles), the sample of \cite{2012ApJ...754...98B} (black filled circles). The solid black line represents a linear fit of the local MZR derived by \cite{2012ApJ...754...98B}, with the shaded gray region indicating the $1\sigma$ intrinsic scatter. The solid red curve represents the MZR calibration for $z\sim$ 0 from \cite{2013ApJ...765..140A}. The horizontal dashed line marks the locus of the extremely metal-deficient \citep[XMD; metallicity $\lesssim$ 10\% Z$_{\odot}$, see][]{2000A&ARv..10....1K}, highlighting that most of our EEMPGs have extremely low metal abundances.}
  \label{MZR}
\end{figure} 

\section{Extreme emission line metal-poor galaxy sample} \label{Sec_sample}

\subsection{Selected galaxies}

Our sample is constructed as follows: five galaxies of our sample were observed on the Cycle 2 JWST program PID 3226 (PI: Schaerer). These galaxies were selected from a comprehensive sample identified by \cite{2021A&A...646A.138I} using the Sloan Digital Sky Survey \citep[SDSS;][]{2000AJ....120.1579Y}, based on the following primary criteria: high equivalent widths of EW(\Hb) $\geq$ 100 $\AA$, which suggest a recent starburst, elevated sSFR, and efficient production of ionizing photons, and oxygen abundances 12 + log(O/H) $\leq$ 7.25 \citep[direct $T_e$-method, ][]{2015MNRAS.451.2251I}. A compactness criterion was also applied during observations of these galaxies with the Hubble telescope. The other six EEMPGs originate from the Cycle 1 JWST program PID 2424 (PI: Jaskot). In Table \ref{table:sample_properties}, we present our galaxy sample along with their physical properties, such as their redshift, oxygen abundance, stellar mass, and star formation rates, and equivalent widths of H$\beta$. 
The stellar masses have been derived using the methods described in \cite{2021A&A...646A.138I}.

\begin{table*}[t]
\caption{Coordinates, redshifts, gas-phase metallicities, stellar masses, SFRs, and EWs of H$\beta$ of the sample galaxies.}
\label{table:sample_properties}
\centering
\begin{tabular}{l r r c c c c c c}
\hline\hline
Object & \multicolumn{1}{c}{RA} & \multicolumn{1}{c}{Dec.} & $z$ & $12+\log(\text{O/H})$ & \multicolumn{1}{c}{$\log M_*$} & \multicolumn{1}{c}{$\text{SFR(H}\beta)$} & EW(H$\beta$)\\
 & \multicolumn{1}{c}{(deg)} & \multicolumn{1}{c}{(deg)} & & & \multicolumn{1}{c}{$(M_\odot)$} & \multicolumn{1}{c}{$(M_\odot\,\text{yr}^{-1})$} & \multicolumn{1}{c}{($\AA$)} \\
\hline
J0811+4730      & 122.967 & 47.507 & 0.04447 & 6.97 & 5.79 & 0.193 & 318.7 \\
J1004+3256      & 151.041 & 32.936 & 0.06622 & 7.16 & 6.13 & 0.400 & 459.0 \\
J104458+03531   & 161.240 &  3.886 & 0.01287 & 7.44 & 7.83 & 0.170 & 106.8 \\
J1155+5739      & 178.868 & 57.664 & 0.01714 & 7.95 & 7.18 & 0.989 & 169.0 \\
J120202+54155   & 180.510 & 54.264 & 0.01203 & 7.48 & 6.10 & 0.136 & 254.0 \\ 
J122437+37243   & 186.153 & 37.410 & 0.04058 & 7.82 & 7.97 & 1.059 & 115.8 \\
J1234+3901      & 188.565 & 39.021 & 0.13310 & 7.03 & 7.93 & 1.567 & 276.0 \\
J1505+3721      & 226.285 & 37.361 & 0.07547 & 7.23 & 6.47 & 0.640 & 298.0 \\
J150934+37314   & 227.392 & 37.529 & 0.03261 & 7.85 & 7.07 & 1.799 & 237.3 \\
J160810+35280   & 242.043 & 35.469 & 0.03275 & 7.82 & 7.04 & 0.596 & 342.3 \\
J2229+2725      & 337.387 & 27.423 & 0.07611 & 7.09 & 6.96 & 0.680 & 577.0 \\
\hline
\end{tabular}
\tablefoot{Redshifts taken from \cite{2024MNRAS.527..281I} and SDSS DR18, oxygen abundance $12+\log(\text{O/H})$, stellar mass, SFRs and EWs are obtained following \cite{2021A&A...646A.138I}.}
\end{table*}


\subsection{Fundamental galaxy properties of our EEMPGs}

Galaxy evolution is characterized by strong correlations between fundamental properties such as stellar mass, star formation rate, and metallicity. These global parameters of our EEMPG sample, obtained from the optical data, are presented in Table \ref{table:sample_properties}. This section outlines the characteristics and positions of our EEMPGs on diagrams of fundamental attributes like the galaxy main sequence (MS) and mass-metallicity relation (MZR), comparing them to other relevant samples. These comparison samples include a local dwarf galaxy sample from \cite{2012ApJ...754...98B} originating from the \textit{Spitzer} Local Volume Legacy (LVL) survey \citep{2009ApJ...703..517D}, the CSFG \citep{2021A&A...646A.138I}, a local sample of typical MS SFGs \citep[HECATE;][]{2026MNRAS.548ag522K}, and two prototypical metal-poor starburst galaxies, I Zw 18 and SBS 0335-052 E \citep{1997ApJ...476..698I,2025ApJ...985..253M}.



Star-forming galaxies exhibit a strong correlation between their star formation rate (SFR) and stellar mass ($M_*$) which is often described as the MS of galaxies and has been studied by many works in the past \cite[e.g.,][]{2007A&A...468...33E,2007ApJ...670..156D,2007ApJ...660L..43N,2014ApJS..214...15S,2017MNRAS.466.1192M}. This relation suggests that typical galaxies experience a steady growth regulated by gas accretion and feedback \citep[e.g.,][]{2012ApJ...754L..29W}. Deviations above this relation, such as higher SFRs for a given $M_*$, imply increased star-formation efficiencies driven by stochastic processes like mergers or the rapid inflow of pristine gas \citep{2014ApJS..214...15S}. Studying extreme galaxies, such as those with metal deficiency and/or with intense starburst activity, within this framework allows for quantifying their evolution relative to the typical population.

Figure \ref{MS} shows the location of our EEMPG sample in respect to two local samples of star-forming galaxies; the first is taken from the second release of Heraklion Extragalactic Catalog\footnote{\url{https://hecate.ia.forth.gr}} \citep[HECATE;][]{2021MNRAS.506.1896K,2026MNRAS.548ag522K}, a value-added catalog for multi-wavelength surveys in the local Universe and the second is from \cite{2021A&A...646A.138I} which is a local sample of compact star-forming galaxies (CSFGs). In Fig. \ref{MS} we see that our EEMPGs are located well above the MS SFGs and in the low-mass tail of the distribution of the CSFGs suggesting that EEMPGs have much higher SFRs per unit $M_*$ (i.e., specific star formation rates, sSFRs) compared to the typical SFG and CSFG populations, which is expected given our selection criteria. Figure \ref{MS} also shows that our EEMPGs have SFRs intermediate between the two starbursts and generally similar $M_*$, though nearly half of our galaxies are approximately one order of magnitude more massive.

The stellar mass of a galaxy is fundamentally linked to its gas-phase metallicity, a correlation known as the mass-metallicity relation (MZR) \citep[e.g.,][]{2004ApJ...613..898T, 2013ApJ...779..102K, 2023ApJS..269...33N}. In Figure \ref{MZR}, we present our EEMPG sample with other similar populations, including the CSFGs from \cite{2021A&A...646A.138I} and a representative local sample of dwarf galaxies from \cite{2012ApJ...754...98B}. While our EEMPGs generally align with the linear MZR derived by \cite{2012ApJ...754...98B}. Nonetheless, there are two distinct outliers in our EEMPG sample that significantly deviate from both fits (having 12+log(O/H) < 7.25 and $M_*>6.9~M_{\odot}$), which is anticipated for metal-poor galaxies \citep[e.g., see][]{2024ApJ...971...43M,2020ApJ...891..181M} and seem to be almost randomly distributed over a wide range of stellar masses \citep{2025ApJ...991..191B}.

\section{JWST observations} \label{Sec_JWST_obs}

\subsection{Data reduction and analysis}

The uncalibrated data of our sample (see Table \ref{table:sample_properties}) were processed using JWST Science Calibration Pipeline v1.17.1, producing twelve rectified cubes per target combined via the drizzle algorithm \citep{2023AJ....166...45L}. The detailed reduction steps are described in Appendix\,\ref{JWST_pipeline}. In summary, the pipeline performed detector fringe correction, flux-calibration, and dedicated background exposures were subtracted. We then refined the source centroids in each channel, especially for program 3226 whose targets are relatively faint. Sources were then classified as being point-like, compact, or extended. For point and compact sources, we used a 2D optimal extraction method for each wavelength slice to achieve the best S/N possible. For extended sources we integrated the flux using an aperture driven by the enclosed energy in the brightest lines. This produced a set of spectra from which measure the emission lines.


\subsection{Observed emission lines in mid-IR} 

The JWST/MIRI (medium resolution) spectra of our EEMPG sample range from 5--25\,$\mu$m. The motivation for using mid-IR emission features is that while standard optical emission lines probe a relatively narrow energy range (I.P. $\sim$13.6--40 eV), the mid-IR fine-structure lines of $\mathrm{Ne}$, $\mathrm{Ar}$, $\mathrm{S}$, $\mathrm{O}$, and $\mathrm{Si}$ probe the ionization of the interstellar medium (ISM) to much higher energies (e.g., \Nev\ 14.3\,$\mu$m, I.P. 96 eV). Measuring this wide range of ionization stages is essential to resolve their dominant source of ionization. Specifically, high-excitation lines like \Nev\ 14.3\,$\mu$m directly constrain the extreme ultraviolet (EUV) and soft X-ray source spectral energy distribution (SED) at energies >54 eV, whereas low-ionization lines such as \Arii\ 6.98\,$\mu$m and \Neii\ 12.81\,$\mu$m are sensitively boosted by X-ray photoionization. Simultaneously measuring these features allows us to distinguish between competing high-energy stellar and non-stellar sources, including binary stars, shocks, and ultraluminous X-ray sources (ULXs). Selecting this wavelength range targets significant emission lines such as the main hydrogen recombination lines ($\mathrm{Pf}\alpha$, $\mathrm{Hu}\alpha$, and $\mathrm{Hu}\beta$), as well as the low-to-intermediate ionization fine-structure lines, including \Siv\ 10.51\,$\mu$m, \Neii\ 12.81\,$\mu$m, \Nev\ 14.32\,$\mu$m, \Neiii\ 15.56\,$\mu$m, \Siii\ 18.71\,$\mu$m, and \Oiv\ 25.89\,$\mu$m (see Table \ref{table:line_list}). 

In this analysis, we consider a measurement as a detection if its signal-to-noise ratio (S/N) exceeds 3. For our EEMPG sample, we include non-detections (S/N < 3) as upper limits, using in general 3$\sigma$ upper limits. For flux ratio calculations with unreliable measurements, non-detections in the numerator or denominator are replaced by 3$\sigma$ upper limits, resulting in upper or lower limits on ratios. If both numerator and denominator are undetected, we do not report any value.

From our measurements, we find three objects with \Nev\ 14.32\,$\mu$m (I.P. 97 eV) detections, and four galaxies with \Oiv\ (I.P. 54 eV). In total, 5 out of 11 EEMPGs show the presence of mid-IR lines which we consider as ``high ionization'' lines, i.e.~with I.P. $>54$ eV. Multiple other fine structure lines as well as H recombination lines are also detected in these objects. The complete list of our targeted emission lines is presented in Appendix \ref{apdx_obslnrts} (Table \ref{table:calculated_ratios} and \ref{table:calculated_fluxes}).

\begin{figure}[htb]
  \resizebox{\hsize}{!}{\includegraphics{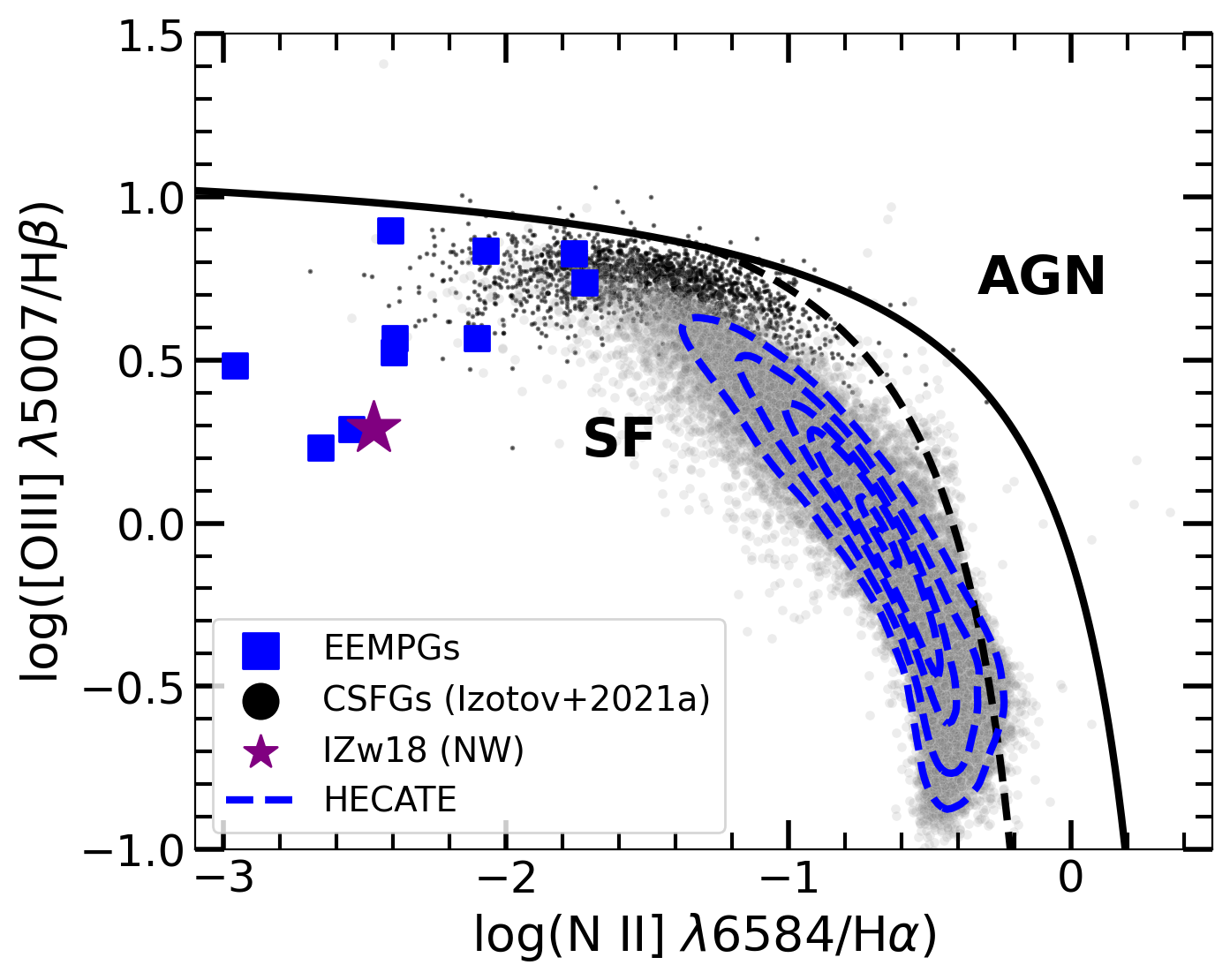}}
  \caption{\Oiii$\lambda5007$/H$\beta$ against the \Nii $\lambda6584$/H$\alpha$ for various star forming galaxy populations; CSFGs \citep{2021A&A...646A.138I} (black points), and the extremely metal-deficient galaxy I Zw 18 (NW) (purple star) \citep{1998ApJ...497..227I}. Blue dashed contours (and grey points) represent the density distribution of the HECATE galaxy sample, representing typical local star-forming galaxies. Our EEMPGs (blue squares) are located primarily in the upper-left quadrant, characteristic of low-metallicity, high-excitation star-forming regions. The solid black line represents the maximum theoretical starburst line \citep{2001ApJ...556..121K} and the dashed black line represents the empirical demarcation line between star-forming and AGN galaxies \citep{2003MNRAS.346.1055K}. SBS 0335-052E is excluded from this plot due to the absence of literature data for the \Nii$\lambda$6584 line.}
  \label{BPT-OIII}
\end{figure}

\section{Emission-line diagnostics of EEMPGs} \label{Sec_MIR_diagnostics}

Analyzing emission lines offers significant insights into a galaxy's activity, the conditions in its ISM, and its ionization fields. Consequently, the study of emission-line diagnostics enables the examination of specific conditions within a galaxy. To contextualize our EEMPG sample within the broader galaxy populations, we incorporate comparison samples of blue compact dwarfs (BCDs), local samples of typical and compact (with higher sSFRs) star-forming galaxies, luminous infrared galaxies (LIRGs), and AGNs. This includes in particular the two prototypical metal-poor galaxies, I Zw 18 \citep{1972ApJ...173...25S,1999ApJ...511..639I,2025ApJ...992...48H} and \sbs\ \citep{1990Natur.343..238I,2025ApJ...985..253M}, which were recently observed with MIRI. 

\subsection{Optical diagnostic diagrams}

Optical emission line ratios have been extensively utilized as diagnostic tools for assessing activity, as they are sensitive to the ionization mechanisms and the physical conditions present within the gas. Among the most commonly utilized diagrams is the \Oiii$\lambda$5007/\Hb\ versus \Nii $\lambda$6584/H$\alpha$, introduced by Baldwin-Phillips-Terlevich \citep[hereafter BPT diagram;][]{1981PASP...93....5B}. This diagram features demarcation lines designed to distinguish pure star-forming galaxies from AGN-dominated galaxies \citep{2001ApJ...556..121K,2003MNRAS.346.1055K}.

Figure \ref{BPT-OIII} shows the location of our EEMPG sample in relation to the CSFGs sample \citep{2021A&A...646A.138I} and a sample of local (typical) SFGs \citep[HECATE;][]{2026MNRAS.548ag522K}. Our EEMPGs are located in the region of star-forming (SF) galaxies. Approximately half of the objects are closely clustered near the SF-AGN boundary as defined by \cite{2001ApJ...556..121K}, indicating the presence of intense ionization fields and/or low metallicities \citep{2013ApJ...774..100K}. The remaining objects are situated in proximity to the position of I Zw 18. The entire sample demonstrates lower \Nii $\lambda$6584/H$\alpha$ ratios compared to the CSFGs (our sample exhibits values below -1.6), which is attributable to their extremely low metallicities.

Another widely used emission-line ratio diagnostic for the ionization conditions within star-forming galaxies is O32 = \Oiii$\lambda$5007/\Oii$\lambda$3727. Since this ratio compares emission from two different ionization stages of the same element, it primarily traces the ionization parameter and the hardness of the radiation field produced by young massive stars. Compact star-forming regions found in low-metallicity environments are typically linked to elevated O32 values, indicating intense recent star formation. The ratio is also sensitive to the shape of the ionizing spectrum and is therefore commonly employed in studies of extreme emission-line galaxies \citep{2014MNRAS.442..900N,2019MNRAS.489.2572T,2024MNRAS.535.1796B}. 

\begin{figure}[htb]
  \resizebox{\hsize}{!}{\includegraphics{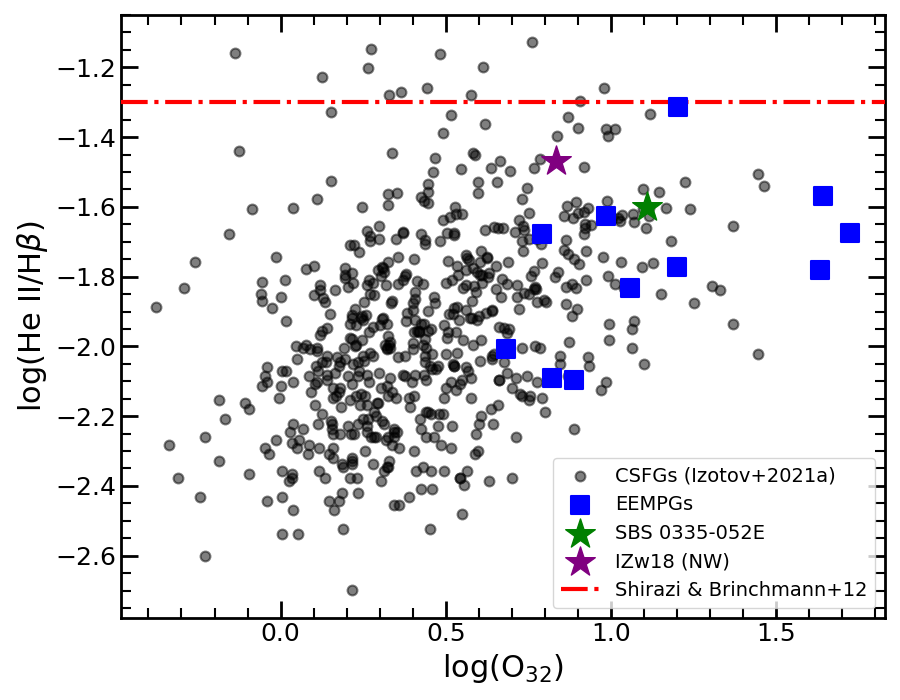}}
  \caption{Nebular line ratios \Heii\,$\lambda4686$/\Hb\ versus O32 of our EEMPGs and comparison samples. The EEMPGs (blue squares) occupy systematically higher O32 and \Heii\,$\lambda4686$/\Hb\ values compared to the reference CSFGs (black disks), indicating harder radiation fields and harder ionized nebular conditions. The location of our EEMPGs near the limit separating star-forming galaxies from AGNs \citep{2012MNRAS.421.1043S} suggests that the emission from our galaxies may not be explained by standard single stellar populations alone.
  }
  \label{O32_HeII_Hb}
\end{figure}

Nebular \Heiiopt\ emission is present in all of our EEMPGs. The \Heiiopt/\Hb\ ratio is an important diagnostic of the hardness of the ionizing radiation field in galaxies. Since the production of nebular \Heii\ emission requires photons with energies above 54.4 eV, significantly higher than the \Oiii\,$\lambda$5007 (35.1 eV), this ratio traces the presence of very energetic ionizing sources. Elevated \Heii\,$\lambda4686$/\Hb\ values are commonly associated with young low-metallicity stellar populations, Wolf-Rayet (WR) stars, X-ray binaries, fast radiative shocks, or AGN. In star-forming galaxies, the ratio is often used to probe extreme ionization conditions and to constrain the contribution of hard ionizing spectra beyond that produced by standard stellar populations \citep{2015ApJ...801L..28K,2012MNRAS.421.1043S}.

Figure \ref{O32_HeII_Hb} shows that our EEMPGs have generally high values of both O32 and \Heiiopt/\Hb\ ratios. Furthermore, we see that our EEMPGs are located below but relatively close to the demarcation line that separates AGN from extreme starbursts defined by \cite{2012MNRAS.421.1043S}.  This line establishes the theoretical upper limit, which is largely insensitive to metallicity, of \Heiiopt/\Hb\ emission achievable by a purely stellar population, effectively isolating extreme starbursts from AGN-dominated galaxies. Specifically, since this limit defines a locus where 10\% of the galaxy's \Heiiopt/\Hb\ flux is contributed by an AGN, this does not exclude the possibility that our EEMPGs may have a non-negligible contribution from an AGN aiding to the hardness of the radiation field.

\subsection{Infrared diagnostic diagrams}

\begin{figure*}[h!]
  \resizebox{\hsize}{!}{\includegraphics{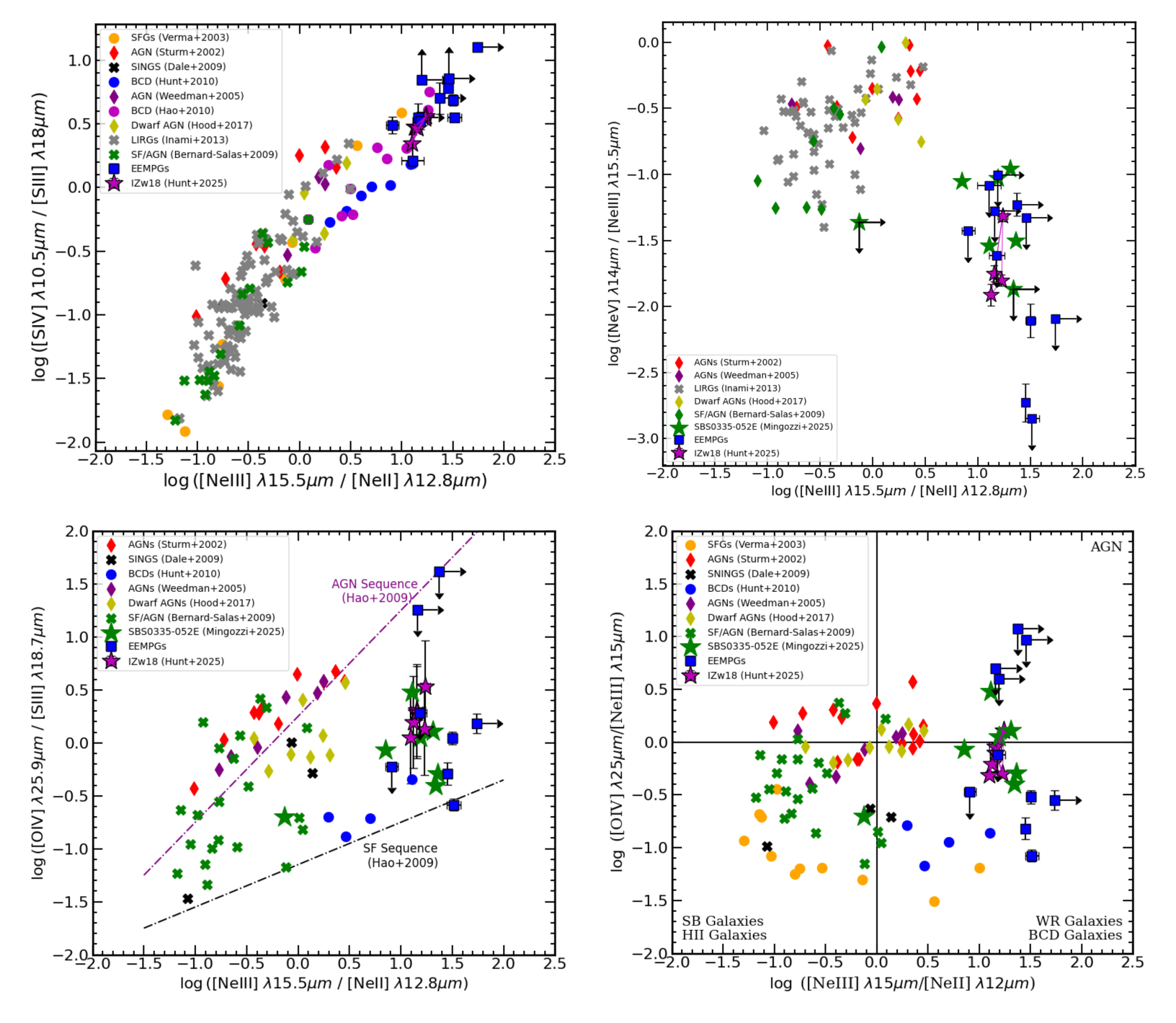}}
  \caption{Mid-IR emission line flux ratio plots of log(\Siv/\Siii) (top left), log(\Nev/\Neiii) (top right), log(\Oiv/\Siii) (bottom left), and log(\Oiv/\Neiii) (bottom right) against the radiation field hardness ratio of log(\Neiii/\Neii). Our EEMPG sample (blue squares) is plotted along with galaxy samples of various types of activity from the literature. Star-forming galaxies \citep[SINGS][]{2009ApJ...693.1821D,2003A&A...403..829V}, AGNs \citep{2002A&A...393..821S,2005ApJ...633..706W}, dwarf AGNs \citep{2017ApJ...838...26H}, blue compact dwarfs (BCDs) \citep{2010ApJ...712..164H,2009ApJ...704.1159H}, composite SF+AGN systems \citep{2009ApJS..184..230B}, luminous infrared galaxies \citep[LIRGs;][]{2013ApJ...777..156I}, and the extremely metal-poor galaxy \sbs\ \citep{2025ApJ...985..253M}. I Zw 18 \citep{2025ApJ...992...48H} is included as a reference for an extremely metal-poor galaxy. In the lower-left panel, the star-forming and AGN mixing sequences from \cite{2009ApJ...704.1159H} are overplotted. The lower-right panel includes the commonly adopted diagnostic boundaries separating \ion{H}{II}/starburst galaxies, AGNs, and Wolf-Rayet/BCD galaxies \citep{2010ApJ...716.1151W}. Upper and lower limits are indicated by arrows, while error bars represent the measurement uncertainties. Our EEMPGs occupy the high-ionization loci of these diagnostics, exhibiting elevated \Neiii/\Neii\ ratios compared to normal star-forming galaxies and overlapping primarily with BCDs and other metal-poor systems, indicating the presence of hard radiation fields that include galaxy activity types.}
  \label{IR-diganostics-master}
\end{figure*}

The mid-infrared spectra of galaxies feature a wealth of emission lines generated by diverse ionization fields. Consequently, analyzing specific emission-line ratios allows us to probe the hardness of the underlying ionization field. For this purpose, we utilize well-established diagnostics from the literature \citep[see][]{2009ApJ...704.1159H,2010ApJ...716.1151W,2013ApJ...777..156I,2022ApJ...927..165R,2025ApJ...993..154R} tailored to target both high and low ionization and line ratios (see Table \ref{table:line_list}).

Figure \ref{IR-diganostics-master} shows the location of our EEMPGs on widely used IR activity diagnostics within representative samples of various galaxy activity types for context. These samples include: metal-poor ($12 + \log(\text{O/H})\sim1-50\% Z_{\odot}$) BCDs \citep{2009ApJ...704.1159H,2010ApJ...712..164H}, star-forming galaxies \citep{2003A&A...403..829V,2008ApJ...677..926S,2009ApJS..184..230B,2009ApJ...693.1821D}, LIRGs \citep{2013ApJ...777..156I}, AGNs \citep{2003A&A...403..829V}, and dwarfs galaxies hosting an AGN with black hole masses of $M_{\mathrm{BH}} \lesssim 10^6 M_{\odot}$ \citep{2017ApJ...838...26H} which fall in the regime of the intermediate black holes \citep[IMBHs ($10^3 \lesssim M_{\mathrm{BH}} \lesssim 10^6 M_{\odot})$;][]{2024MNRAS.531.4311B}. 
%

\begin{table}[t]
\caption{List of our adopted mid-IR diagnostic emission lines, their emitting species, energy ranges of existence, and wavelengths.}
\label{table:line_list}
\centering
\begin{tabular*}{\columnwidth}{@{\extracolsep{\fill}}llcc}
\hline\hline
Line & Species & Energy Range & $\lambda$ \\
     &         & (eV)         & ($\mu$m) \\
\hline
\Neii            & Ne$^+$    & 21.56 -- 40.96    & 12.81  \\
\Siii            & S$^{2+}$  & 23.34 -- 34.79    & 18.71  \\
\Siv             & S$^{3+}$  & 34.79 -- 47.30    & 10.51  \\
\Neiii           & Ne$^{2+}$ & 40.96 -- 63.45    & 15.56  \\
\Oiv             & O$^{3+}$  & 54.95 -- 77.41    & 25.89  \\
\Nev             & Ne$^{4+}$ & 97.12 -- 126.21   & 14.32  \\
\hline
\end{tabular*}
\tablefoot{The lower bound of the energy range is the ionization potential of each species and the upper bound is the energy required to destroy it by ionizing it to the next higher state. Data taken from National Institute of Standards and Technology (NIST)\footnote{\url{https://www.nist.gov/pml/atomic-spectra-database}} Atomic Spectra Database.}
\end{table}

Typically, the strongest mid-IR lines for which the largest number of measurements are available include \Neiii, \Neii, \Siv, and \Siii. Starting with the flux ratio of \Siv/\Siii\ versus \Neiii/\Neii\ presented on the top left panel of Fig. \ref{IR-diganostics-master}. The \Siv/\Siii\ and \Neiii/\Neii\ ratios trace the hardness of the radiation field as their ionization potentials differ by $\sim$10 and $\sim$20 eV, respectively \citep[e.g.,][]{2000ApJ...539..641T,2002A&A...393..821S,2003A&A...403..829V}. In this plot, we see a clear correlation for all samples between the two line ratios. Our EEMPGs are located in the most extreme high-ionization end of this trend, even surpassing I Zw 18, suggesting the presence of the most extreme ionization field found in metal-poor galaxies so far (see also Fig. \ref{O32_HeII_Hb}). 

The \Nev/\Neiii\ against \Neiii/\Neii\ is presented in the top right panel of Fig. \ref{IR-diganostics-master}. In that plot, we see the star-forming, AGNs, and dwarf galaxy samples from the literature, largely populating the top-left to central region of the plot characterized by moderate to low ionization states (\Neiii/\Neii) $\lesssim 3$) and a wide spread of relatively robust \Nev\ emission. Conversely, our EEMPGs are uniquely offset toward the bottom-right. These sources exhibit extreme ionization fields, traced by highly elevated \Neiii/\Neii\ ratios (generally $>10$), while simultaneously showing lower and strongly suppressed \Nev/\Neiii\ values or upper limits. A broad, inverse sequence connects these samples, with local BCDs spanning the intermediate space. Notably, \sbs\ and I Zw 18 are located to the left of our EEMPGs, suggesting that our EEMPG galaxies are extreme, highly ionized environments. These results suggest that our EEMPGs have more prominent emission in \Neiii\ than the rest of the extreme emission line galaxies (BCDs or AGNs), lowering the \Nev/\Neiii\ and enhancing \Neiii/\Neii.


In the bottom left panel of Fig. \ref{IR-diganostics-master}, we see the \Oiv/\Siii\ ratio against the \Neiii/\Neii\ ratio, along with the star-forming and AGN sequences defined by \cite{2009ApJ...704.1159H}. Our EEMPGs exhibit \Neiii/\Neii\ ratios $ > 1$, positioning them at the far right of the SF sequence. This shift toward higher values indicates that our EEMPGs have a higher ionization parameter ($U$) and a harder ionizing radiation field compared to typical SF galaxies, reflecting an elevated excitation state of the gas \citep{2003A&A...403..829V,2019ARA&A..57..511K}. Additionally, the \Oiv/\Siii\ ratio places our EEMPG sample between the sequences of star-forming and AGN galaxies \citep{2009ApJ...704.1159H}, among the \sbs\ and I Zw 18.

The \Oiv/\Neiii\ against \Neiii/\Neii\ diagram presented on the bottom right panel of Fig. \ref{IR-diganostics-master}, has been proposed by \cite{2010ApJ...716.1151W} as a separation of star-forming, BCDs, and AGN galaxies. In this plot, half of our EEMPGs are situated in the quadrant typically associated with BCDs, while the other half, albeit with upper limits, are found in the AGN quadrant. Notably, all of our EEMPGs with higher \Oiv/\Neiii\ ratios than the rest of the BCDs (in the AGN locus) have limits for the \Oiv/\Neiii\ ratio, so we cannot rule out that these objects are compatible with being WR galaxies or BCDs.

\section{Hardness of the ionizing radiation field and possible sources of high ionization lines} \label{Sec_Origin_of_high_IP_lines}

\subsection{\Heii\ $\lambda$4686\AA\ and \Nev\ 14.3\,$\mu$m detections}

\Heiiopt\ is detected in all the objects of our EEMPG sample, with relative intensities $\sim$1--4\% of the \Hb\ flux, and our observations show three new mid-IR \Nev\ line detections (in J104458+03531, J1505+3721, and J150934+37314). Among the EEMPG sample observed with MIRI so far, this brings the total of mid-IR \Nev\ detections to 6 out of 14 (our EEMPGs and including \sbs, I Zw 18, and CGCG007-025), thus demonstrating the presence of ionizing radiation extending beyond $\ga 97.1$ eV in $\sim 42$ \% in the local EEMPGs.

\begin{figure}[tb]
  \resizebox{\hsize}{!}{\includegraphics{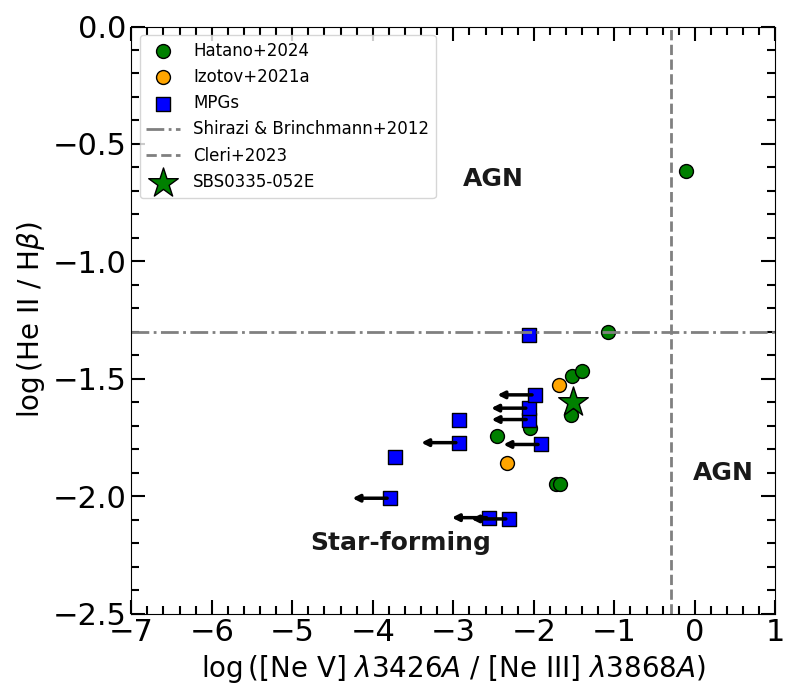}}
  \caption{Diagnostic diagram of log(\Heii\ $\lambda$4686/\Hb) versus log(\Nev $\lambda$3426/\Neiii $\lambda$3868) introduced by \cite{2024MNRAS.534.2633C}, comparing our observations with samples of dwarf galaxies with strong high-ionization lines from the literature (\cite{2024ApJ...966..170H}; green circles and \cite{2021A&A...646A.138I}; orange circles) and the \sbs\ \citep[green star;][]{2005ApJS..161..240T}. According to the analysis of \cite{2024MNRAS.534.2633C}, our EEMPG sample (blue squares) can be explained by a small AGN contribution of 4--8\%. We also show AGN demarcation lines for both \Heii\ $\lambda$4686/\Hb\ \citep{2012MNRAS.421.1043S} and \Nev $\lambda$3426/\Neiii $\lambda$3868 \citep{2023ApJ...953...10C}.}
  \label{HeII_NeVopt_Clsm}
\end{figure}


\cite{2024MNRAS.534.2633C} recently introduced a diagnostic diagram using high-excitation optical emission line ratios of \Heii\ $\lambda$4686/\Hb\ against \Nev $\lambda$3426/\Neiii $\lambda$3868. Since our EEMPG sample lacks observational data for \Nev $\lambda$3426, we use our JWST/MIRI observations of \Nev\ 14.3\,$\mu$m and the \Neiii\ 15.5\,$\mu$m and \texttt{PyNeb} \citep{2015A&A...573A..42L} to predict \Nev $\lambda$3426/\Neiii $\lambda$3868 ratio. In Appendix \ref{PyNeb_for_opt_NeV} we describe the process of predicting these lines and we provide a relation for translating the infrared to optical \Nev/\Neiii\ ratio. Using these predictions we plot in Fig.~\ref{HeII_NeVopt_Clsm} the \Heii\ $\lambda$4686/\Hb\ against \Nev $\lambda$3426/\Neiii $\lambda$3868 for our EEMPG sample. There, we see that it is located among other samples from the literature with strong high-ionization lines \citep{2024ApJ...966..170H,2021A&A...646A.138I} and the \sbs\ \citep{2005ApJS..161..240T}. In particular, our EEMPGs are located outside the AGN locus (below the line of \cite{2012MNRAS.421.1043S} and left from the line of \cite{2023ApJ...953...10C}). However, following the work of \cite{2024MNRAS.534.2633C}, the location of these galaxies in Fig. \ref{HeII_NeVopt_Clsm}, which could be explained by a combination of young stellar populations and IMBH accretion, with the black hole supplying a small fraction (4–8\%) of the total hydrogen-ionizing photon budget.

\subsection{Are high ionization lines from IMBH or shocks ?}

While intense photoionization from stellar populations drives the primary emission spectra of metal-poor dwarf starburst galaxies, pure stellar radiation models routinely struggle to reproduce their high-ionization emission lines (e.g., \Nev\ and the presence of strong \Heii\ \citep{2012MNRAS.421.1043S,2019A&A...621A.105S}. In these low-metallicity starburst environments, radiative feedback from massive stellar winds and core-collapse supernovae naturally drives fast shocks through the interstellar medium, producing the high-energy photons required to reach these extreme ionization \citep[e.g.,][]{2004ApJS..153...75G,2005ApJS..161..240T,2012MNRAS.427.1229I,2021A&A...646A.138I}. However, the presence of extremely young populations should dominate the majority of lower ionization lines ($\lesssim$ 54 eV). Furthermore, pure shock models seem to overpredict lower ionization lines \citep[e.g., \Neii,][]{2025ApJ...985..253M} shifting the relevant ratios (e.g., \Neiii/\Neii) away from the observations. \cite{2012MNRAS.427.1229I} found that stellar sources alone cannot reproduce the observed \Nev$\lambda$3426/\Heii\ $\lambda$4686 ratio in metal-poor dwarf galaxies, but that supplementing the stellar radiation field with just a $\sim$10\% ionizing contribution from radiative shocks is sufficient to explain the high ionization potential lines. Therefore, blending radiative shock models with young simple stellar populations (SSPs) establishes a physically motivated framework to explain all the observed trends in the commonly observed line ratios.

\subsubsection{Models including contributions from shocks}

To test whether radiative shocks drive the observed emission-line ratios in these galaxies, we modeled composite ratios. These were generated by linearly blending predicted line intensities from shock models and young stellar populations, using their relative fractional \Hb\ contributions as the scaling parameter. This follows the assumption that the excitation of each process occurs in different parts of the galaxy. In other words, all individual model line fluxes $I_\lambda$ are normalized to the $\text{H}\beta$ flux, the composite flux for any ratio is calculated as:
\begin{equation}
    I_{\lambda,\text{mix}} = f_{\text{H}\beta} I_{\lambda,\text{shock}} + (1 - f_{\text{H}\beta}) I_{\lambda,\text{SSP}},
    \label{eq:blended_flux}
\end{equation}
where $f_{\text{H}\beta}$ represents the fractional contribution of the shock component to the total $\text{H}\beta$ luminosity of the blended models. In addition, the SSP models account for X-ray binary contribution from stellar evolution \citep{2024ApJ...960...13G}.

We adopt the shock models derived by \citet{2025MNRAS.543.3367F}\footnote{\url{https://zenodo.org/records/17459088}} (precursor+shock) using MAPPINGS-V \citep{2017ApJS..229...34S} and stellar photoionization models which are provided by \cite{2024ApJ...960...13G} based on binary population synthesis models developed by \cite{2013ApJ...764...41F} and BPASS v2.2 \citep{2017PASA...34...58E} with burst ages spanning 1 to 20 Myr to model SSPs, across diverse physical conditions both for the SSP and shock parameters. Specifically, for SSPs we include three values of the ionization parameter (log$U$ = --3.0, --2.0, and --1.0) and four stellar population ages (1, 5, 10, and 20 Myr). For the shock models we use three values for the shock velocity ($v = 150, 250, \rm and ~500$ km s$^{-1}$, see \citealt{2012MNRAS.427.1229I}) and three values for the pre-shock hydrogen number density (log n$_H$ = 0, 1, and 2 cm$^{-3}$). Varying the magnetic field strength of the shock did not show any significant impact on the resulting blended points (shocks+SSP), but higher magnetic field values were more successful in reproducing some of the line ratios, in particular, the \Oiv/\Siii\ and \Siv/\Siii\ for which we observed a shift by 0.2 dex. Therefore, we fixed the magnetic field strength to logB$_o=1~\mu $G \citep[see also][]{2025ApJ...992...48H}. For self-consistency between the SSP, the shock models, and our EEMPG sample we fix the metallicity to $Z = 5\% Z_\odot$. The selection of these parameters covers a broad range of possible conditions and results in 108 blended (mixing) points based on equation \ref{eq:blended_flux}.  

\begin{figure*}[h!]
  \centering
  \includegraphics[scale=0.21]{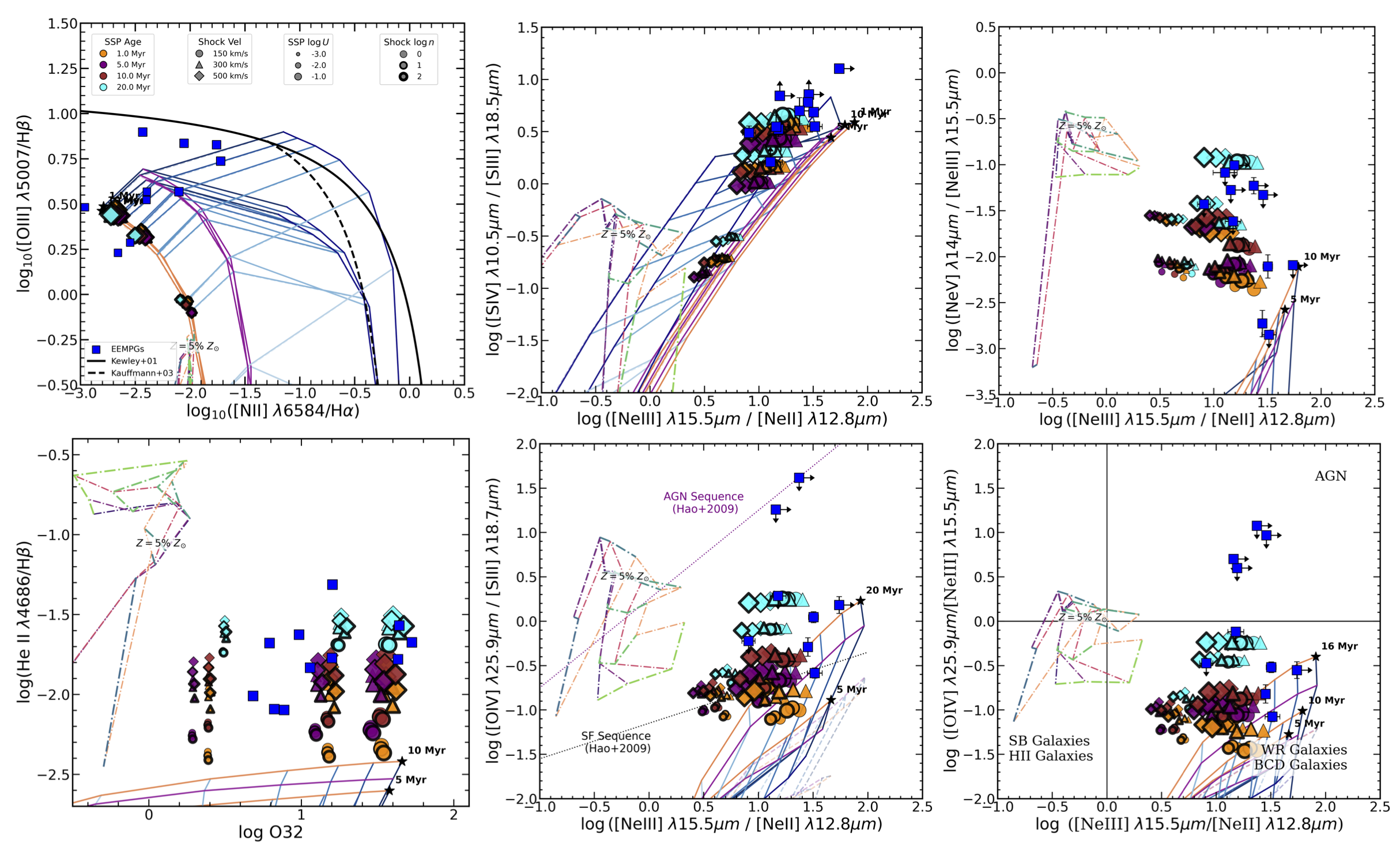} 
  \caption{Optical and mid-IR diagnostic line ratio diagrams comparing our EEMPG sample with stellar photoionization, radiative shock models, and a linear mix of them (blend points of SSP+Shocks). The symbols shown on the panel in the top-left corner for the SSP and Shock models represent the parameter space of each of these models. In all plots, the blending points correspond to $f_{\rm H\beta} = 7\%$, which is the optimal value to better match the location of our EEMPGs in all emission line ratios simultaneously. The first column shows optical emission line ratios. Starting from the top-left panel, we show the BPT diagram \citep{1981PASP...93....5B} of log(\Oiii $\lambda$5007/\Hb) against log(\Nii\ $\lambda$6584/\Ha) with the demarcation lines of maximum star forming activity \citep[][solid line]{2001ApJ...556..121K}, pure star-forming \citep[][dashed line]{2003MNRAS.346.1055K}. The bottom-left panel presents the high-excitation optical diagnostic plotting of log(\Heii\ $\lambda$ 4686/\Hb) against logO32 (where O32 $\equiv$ \Oiii\ $\lambda$ 5007/\Oii\ $\lambda$3727). The remaining four panels (middle and last columns) show the main mid-IR fine-structure line ratios against \Neiii/\Neii, as shown in Fig.~\protect\ref{IR-diganostics-master}. For all blended points (SSP+Shock models) in all plots, the symbol color indicates SSP age (1.0, 5.0, 10.0, and 20.0~Myr) and the symbol size corresponds to the ionization parameter (log$U$ = --3.0, --2.0, and --1.0), as well as radiative shock model grids tracking shock velocities from 150 to 500 km s$^{-1}$ across pre-shock hydrogen densities of log(n/cm$^{-3}$) = 0, 1, and 2. In all plots, we show the grids from shocks (dashed-dotted line) and SSP models (solid and dashed lines) used to sample and create our blended points from shocks \citep{2025MNRAS.543.3367F} and \citep{2024ApJ...960...13G}. Error bars for EEMPGs indicate 1$\sigma$ measurement uncertainties; upper limits correspond to 3$\sigma$ uncertainties.
  }
  \label{SSP_shock_blending}
\end{figure*}

After using the aforementioned procedure to calculate the blended emission line ratios of interest (see Fig. \ref{IR-diganostics-master}, where we present IR-line diagrams), we plot them along with our EEMPG sample on widely used activity diagnostic plots. We show these plots in Fig. \ref{SSP_shock_blending}, which include: optical \Oiii\,$\lambda$5007/\Hb, \Nii\,$\lambda$6584/\Ha, \Heii\,$\lambda$4686/\Hb, \Oiii\,$\lambda$5007/\Oii\,$\lambda$3726, and mid-IR \Siv/\Siii, \Oiv/\Siii, \Oiv/\Neiii, \Nev/\Neiii, and \Neiii/\Neii\ ratios. Trying empirically different \Hb\ fractional contributions (from 0 to 1) for the blend of SSP+Shocks, it was found that values of $f_{\text{H}\beta}$ $\sim$ 5--7\% best match (cluster around) the location of our EEMPGs simultaneously for all the diagnostic plots.

Figure \ref{SSP_shock_blending} also shows that although the SSP+Shock points are broadly consistent with our EEMPG sample, they require high ionization parameter (log$U$ $\sim$ --1.0) and relatively old stellar populations ($\geq$ 10 Myr) across all panels. This finding suggests that the shocks may not be the main source powering the high ionization lines observed in our EEMPGs, since the high sSFR and strong \Hb\ observed in these galaxies imply the dominance of extremely young stellar populations in their spectra. Furthermore, the presence of thermally dominant radio spectra observed in similar EEMPG systems offers independent evidence for young stellar populations possibly devoid of SNe \citep{2026MNRAS.548ag492B}. Specifically, the observed EWs of \Hb\ (> 100~\AA) suggest stellar population ages of $\lesssim$ 5 Myr \citep{1999ApJS..123....3L}. In addition, after inspecting the JWST/MIRI forbidden and hydrogen lines, we did not noticed any line broadening, suggesting that if there are shocks driving fast outflows, they should have velocities $\lesssim$ 200 km s$^{-1}$.


In a recent study of another metal-poor ($\sim$10\%Z$_\odot$) dwarf galaxy, \citep[CGCG 007-025;][]{2026ApJ...998L..45D} using JWST/MIRI observations, the shocks were ruled out as the driving source of the high I.P. lines (e.g., \Nev). Specifically, it was found that even though shocks can be tuned to reproduce high-ionization lines (e.g., \Nev\ and \Oiv), they systematically fail to simultaneously match the observed low-ionization (e.g., \Neii) and high ionization lines.

\begin{figure*}[h!]
  \centering
  \includegraphics[scale=0.67]{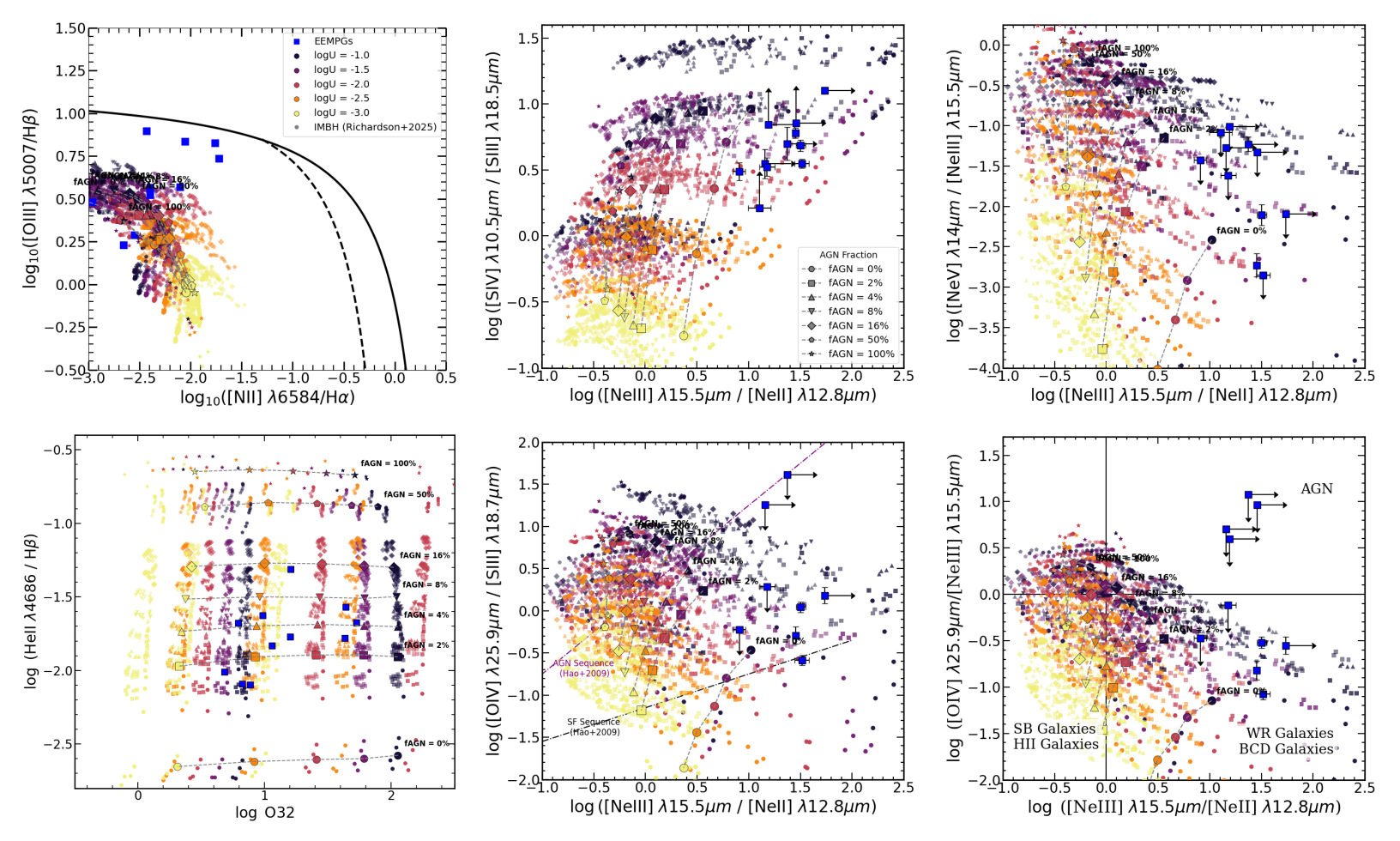} 
  \caption{Diagnostic diagrams, similar to Fig. \ref{SSP_shock_blending} but for the IMBH + SSP models of \cite{2025ApJ...993..154R}. The photoionization model grids are varying in ionization parameter (log$U$ = --1.0 to --3.0; indicated by color scale) and AGN fractional contribution ($f_{\rm AGN}$= 0--100\%) (denoted by symbol type and tracks). Our EEMPG sample is shown in blue squares.}
  \label{IMBH-models}
\end{figure*}

Other studies have investigated the possibility that radiative shocks can explain the presence of high I.P. emission lines in metal-poor environments. Recent works using JWST observations of two prototypical metal-poor starbursts using state-of-the-art fast radiative shocks have shown that such mechanisms fail to reproduce the extreme mid-IR emission line ratios observed in I Zw 18 \citep{2025ApJ...992...48H}. Furthermore, similar analysis of \sbs\ ($\sim$5\%Z$_{\odot}$) by \cite{2025ApJ...985..253M} found that shock models exhibit systematic discrepancies in their predicted mid-IR emission-line strengths. Specifically, low-metallicity shock models tend to substantially overproduce low-ionization lines, most notably \Neii. Consequently, this overproduction results in model predictions that significantly underproduce key emission-line ratios with \Neii\ in the denominator, falling far short of the observed \Nev/\Neii, \Neiii/\Neii, and \Siv/\Neii\ ratios across the galaxy. The shock model grids are restricted to reproducing only the lowest observed limits of \Siv/\Neii\ and \Nev/\Neii, and they cannot account for the highly enhanced, hard high-ionization line ratios such as \Oiv/\Neiii\ or \Nev/\Neiii. While these shock models can match isolated high-ionization line ratios, such as \Oiv/\Neiii, their general inability to replicate the global low-to-high ionization parameter space, combined with the lack of co-spatial [\ion{Fe}{II}] emission, effectively rules out shocks as the dominant mechanism driving the highly excited, hard ionization state of the gas in this local high-redshift analog.

\subsubsection{Models including an IMBH}

Another possible mechanism capable of producing the observed high I.P. lines (i.e., \Nev) in our EEMPGs could be emission from an IMBH. The identification of IMBHs in metal-poor environments is still a challenging task, as metal-poor star-bursting environments complicate standard spectral classification \citep{2021ApJ...906...35S}. Also, traditional AGN photoionization grids from the literature typically struggle in this regime because they do not account for hard ionizing radiation from ultraluminous X-ray sources (ULXs), which penetrate low-metallicity gas and emit hard ionizing radiation that can mimic IMBH excitation. To overcome this degeneracy, we employ the tailored photoionization model suite from \cite{2025ApJ...993..154R}, which uniquely incorporates both IMBHs and evolving ULX populations within a metal-poor, highly star-forming interstellar medium. Furthermore, these models self-consistently parameterize black hole masses based on primordial seeding channels alongside metallicity- and age-dependent ULX population synthesis.

For this analysis, we assume a closed geometry (i.e., spherically symmetric cloud). Although assumptions regarding spatial geometry and gas mixing can significantly alter many simulated emission spectra, \cite{2022ApJ...927..165R} emphasize that key mid-infrared ratios used as activity diagnostics, e.g., \Oiv/\Neiii, are largely unaffected. This insensitivity establishes them as exceptionally reliable diagnostic tools. Furthermore, we limit our grids to a metallicity of 5\%Z$_{\odot}$, and vary the ionization parameter log$U$ of --3 to --1, and AGN fractions from 0 to 100\% (defined as the fraction of the total hydrogen ionizing photons in the composite ionizing continuum that are attributed to the AGN spectral energy distribution). In Fig. \ref{IMBH-models}, we see the location of our EEMPG sample relative to the emission line ratios produced by IMBH+SSP models of \cite{2025ApJ...993..154R}, in the same diagrams as presented in Fig.~\ref{SSP_shock_blending}. There we see emission line ratios covering the spectral range of $\sim$0.4--25\,$\mu$m and ionization potentials from $\sim$20--97\,eV. The panels of Fig.~\ref{IMBH-models} show that the grids are able to self-consistently reproduce the emission line ratios observed across the majority of our objects (excluding upper limits) and across all diagnostic plots, for a similar IMBH fraction and ionization parameter. IMBHs could thus be powering the high ionization lines observed in our EEMPG sample.

JWST/MIRI observations of I Zw 18 \citep{2025ApJ...992...48H} and \sbs\ \citep{2025ApJ...985..253M} show prominent \Oiv\ and \Nev\ emission that can be best modeled with an accreting IMBH. For I Zw 18, photoionization models with a low AGN fraction ($f_{\rm AGN}\sim$~0.04) explain these lines, but an IMBH remains ambiguous because low-metallicity Wolf-Rayet stars or ULXs can also reproduce the line ratios. In \sbs, only photoionization models with a $\sim$10$^5$ M$_\odot$ IMBH and AGN fraction of 4-8\% reproduce the diagnostics, though alternative extreme stellar mechanisms cannot be excluded.

There is growing evidence for the presence of IMBHs in metal-poor starbursts. JWST/MIRI-MRS observations of the metal-poor dwarf galaxy CGCG 007-025 reveal the presence of high ionization mid-IR lines, including \Oiv\ and \Nev\ localized within a compact $\sim$50 pc region. These observed mid-IR line ratios are best reproduced by a hybrid starburst+AGN with the AGN contributing 4\%--8\% to the ionizing budget. This active nucleus is likely powered by an intermediate-mass black hole (M$_{\rm BH}$ = 10$^5$M$_{\odot}$) which is further supported by a prominent co-spatial X-ray source and a broad \Ha\ line component \citep{2026ApJ...998L..45D}. The presence of broad lines in dwarf SF galaxies is a frequent occurrence \citep{2004A&A...415L..27I,2021A&A...646A.138I}. Furthermore, cosmological simulations have shown that hidden black holes ($L_X \lesssim$ 10$^{39}$erg s$^{-1}$) make up to 76\% of all black holes in local dwarf galaxies \citep{2022ApJ...936...82S}.

\subsection{The possibility of ULXs dominating the ionizing spectrum}

ULXs are defined as point-like, non-nuclear objects that are characterized by an extreme X-ray luminosity \citep[$L_\mathrm{X} \sim 10^{39}$ erg s$^{-1}$, see e.g.,][for a review]{2017ARA&A..55..303K}. Their nature is considered to be extreme X-ray binaries powered by gas accreting onto a stellar-mass black hole or neutron star \citep{1999ApJ...519...89C,2014Natur.514..202B}. 

These objects are capable of ionizing helium and producing \Heii\ $\lambda$4686 \citep{2004MNRAS.351L..83K,2019A&A...622L..10S}. Thus, the presence of \Heii\ $\lambda$4686 in our sample may be the result of such objects. In order to test this hypothesis for our EEMPGs, we calculate the observational helium ionizing photon production rate ($Q(\mathrm{He}^+)$), defined as the total number of photons emitted per second by the central source with energies above the $\mathrm{He}^+$ ionization threshold ($h\nu$ > 54.4 eV, or $\lambda < 228$~\AA). $Q(\mathrm{He}^+)$ was derived from the reddening-corrected \Heii $\,\lambda4686$ line luminosity following the Case~B recombination using the equation of \citet{2006agna.book.....O}:
\begin{equation}
    Q(\mathrm{He}^+) = \frac{L(\mathrm{He}^+\,\lambda4686)}{j(\lambda4686)/\alpha_B(\mathrm{He}^+)},
    \label{eq:Q_HeII_general}
\end{equation}
where the effective recombination emissivity denominator, $j(\lambda4686)/\alpha_B(\mathrm{He}^+)$, is evaluated using the quantum mechanical atomic data from \citet{1987MNRAS.224..801H} and \citet{1995MNRAS.272...41S}. Using equation (\ref{eq:Q_HeII_general}) for our EEMPGs, we find that $Q(\mathrm{He}^+)$ $\sim$ 10$^{50-51}$ s$^{-1}$. Typically, a single ULX produces up to $Q(\mathrm{He}^+)\sim$ 10$^{48-49}$ s$^{-1}$\citep[see][and references therein]{2024MNRAS.532.1459G}.

ULXs as the explanation of the \Heiiopt\ detection have also been tested in the past in metal-poor environments. Two notable examples include the I Zw 18 and \sbs\ \citep{2015ApJ...801L..28K,2018MNRAS.480.1081K}. There, the authors concluded that ULXs could not explain the \Heii\ $\lambda$4686 detection, as detailed photoionization modeling showed that high-luminosity X-ray sources are insufficient to explain the observed extensive nebular \Heii\ $\lambda$4686. Specifically, in I Zw 18, a photoionization model of the dominant X-ray binary predicts a \Heii\ $\lambda$4686 luminosity that is approximately two orders of magnitude below the observed luminosity. 
In \sbs, \cite{2018MNRAS.480.1081K} estimate an ionizing photon flux of  $\log Q(\mathrm{He}^+) =36.2$ and 35.4, respectively from the central ULX and the soft diffuse northwest component, which falls drastically short of its total observed nebular photon budget of $Q(\mathrm{He}^+)$ = 3.17 $\times$~10$^{51}$. 

However, although the observed \Heiiopt\ luminosity (and consistently $Q(\mathrm{He}^+)$) of I Zw 18 cannot be explained by the presence of a single ULX, its mid-IR luminosities of forbidden lines are similar to those observed in ULXs. More specifically, the \Nev\ 14.3$\mu$m luminosity of the NW region of I Zw 18 \citep{2025ApJ...992...48H} is comparable to that of Holmberg II X-1 (Oskinova et al., private communication). Furthermore, the \Oiv\ 25.9 $\mu$m luminosity of I Zw 18 is comparable to that of two ULXs, such as NGC 6949-X1 and Holmberg II X-1 \citep{2010ApJ...708..364B,2012ApJ...754...98B}. Although some single ULX could in principe produce the high I.P. mid-IR lines, their inability to account for the \Heiiopt\ emission means they cannot explain the full spectrum. 

Therefore, an IMBH remains the favored mechanism for our EEMPGs galaxies.

\subsection{Contribution of other sources and processes to high-energy ionizing photons}

While SSP models often struggle to reproduce the observed \Heiiopt\ emission in extreme metal-poor galaxies, several alternative stellar and interstellar processes can contribute to the high-energy (> 54.4 eV) photon budget. Notably, standard SSPs typically do not account for extreme stellar evolutionary paths that are favored at low-metallicity environments, such as very massive stars (VMS) \citep[e.g.,][]{2025A&A...698A.262M,2026A&A...711L...3M}, rapidly rotating stars \citep[e.g.,][]{2014ApJS..212...14L,2023A&A...679A.137M}, and binary-stripped helium stars \citep[e.g.,][]{2017PASA...34...58E,2019A&A...629A.134G}. Beyond direct stellar continuum emission, the interstellar medium itself can be a major source of the ionizing radiation. The collective mechanical feedback from stellar winds and supernovae creates superbubbles filled with hot, shock-heated plasma (T $\sim10^{6}$ K). This diffuse hot gas emits heavily in the soft X-ray and extreme ultraviolet regimes. As demonstrated by \cite{2022A&A...661A..67O}, the emission from this hot gas can provide a substantial fraction of the He$^+$ ionizing photons in young starbursts. Finally, it is highly possible, given their extreme nature (high sSFRs), that the high ionization lines observed in these extreme emission line galaxies are not driven by a single dominant source, but rather that multiple mechanisms contribute simultaneously.

\section{Conclusions} \label{Sec_conclusions}

In this work, we presented JWST/MIRI observations of one of the most extreme emission line and metal-poor samples of galaxies (EEMPGs) ever observed. In particular, we studied their properties in comparison to standard galaxy populations, such as other prototypical metal-poor dwarf galaxies, extreme and typical main-sequence star-forming galaxies, as well as various types of AGN galaxies. Finally, we proposed some hypotheses about the source that may drive the extreme I.P. emission lines observed in their optical and infrared spectra. Our main results are summarized as follows: 

\begin{enumerate}
    \item Our EEMPG sample spans a wide range of sSFRs, $M_*$, and metallicities (approximately 2 orders of magnitude). This diversity makes it ideal for studying this galaxy population. Furthermore, it contains some extremely metal-poor objects with hard ionization fields, good analogs for studying the conditions in the first galaxies in high-$z$.
    \item In extreme emission line EEMPGs, the \Nev\ 14.3$\mu$m is present in $\sim$42\% (or 6 out of 14, including I Zw 18, CGCG 007-025, and \sbs) of them, indicating that this is a frequent occurrence.
    \item The most plausible explanation for the high I.P. lines detected in our EEMPG sample seems to be the presence of an IMBH with a low AGN fractional contribution (4--8\%). This small AGN contribution to the ionizing photon budget is enough to produce \Nev\ 14.3\,$\mu$m and simultaneously explain the lower ionization lines from the optical to mid-IR. 
    \item Pure shocks struggle to explain the mid-IR emission line ratios in extreme metal-poor galaxies, primarily because shock models systematically overpredict the strength of low-ionization lines. Furthermore, by calculating the ratio of the required shock power to the available stellar power indicates that stellar feedback is theoretically insufficient to power the observed emission, requiring a more energetic driver.
    \item While composite SSP+shock models broadly match our EEMPGs, they require stellar ages ($\geq$ 10~Myr) that contradict the extremely young populations ($\lesssim$ 5~Myr) implied by the high specific SFRs and strong \Hb\ equivalent widths ($>$ 100~\AA).
    \item In our EEMPG sample, the \Heiiopt\ ionizing photon production rate, $Q(\mathrm{He}^+)$, cannot be explained by solely invoking ULXs, as they show a deficit of $\sim$1-2 orders of magnitude.
\end{enumerate}

\noindent In the future, we plan to further constrain the source of the hard ionization fields in EEMPGs galaxies. This will be achieved by performing a multi-wavelength study spanning from X-ray to the radio regime and by employing statistically robust methods to fit and interpret the observed emission lines using state-of-the-art models and codes currently under development.

\begin{acknowledgements} 
This study is supported by a joint Swiss SNF and French ANR grant ("SpeXion", project number 10002276).
This  work is based on observations made with the NASA/ESA/ CSA James Webb Space Telescope. LR gratefully acknowledges funding from the DFG through an Emmy Noether Research Group (grant number CH2137/1-1). OB was supported by the National Science Foundation under Cooperative Agreement 2421782 and the Simons Foundation grant MPS-AI-00010515 awarded to the NSF-Simons AI Institute for Cosmic Origins - CosmicAI, https://www.cosmicai.org/. YI and NG acknowledges support from the National Academy of Sciences of Ukraine by its project no. 0126U000353
and project No. 224866 supported in the result of the Joint Call ‘Ukrainian-Swiss Joint Research Projects: Call for Proposals 2023’. 
AG acknowledges financial support from the European Union’s Horizon Europe research and innovation programme under the Marie Skłodowska-Curie grant agreement NO HORIZON-MSCA-2025-PF-01-01 101282442. 
The work of SD was supported by the National Science Foundation Graduate Research Fellowship Program under Grant Number DGE-2236637. Any opinions, findings, and conclusions or recommendations expressed in this material are those of the author(s) and do not necessarily reflect the views of the National Science Foundation.
\end{acknowledgements}

%

\bibliographystyle{aa}
\bibliography{references}







   
  



\begin{appendix}

\section{Pipeline steps and data reduction} \label{JWST_pipeline}

\subsection{Detector-Level Calibration and Cube Reconstruction}

The Medium-Resolution Spectrometer (MRS) of the Mid-InfraRed Instrument (MIRI) onboard the James Webb Space Telescope (JWST) covers a spectral range of 4.90--28.70~$\mu$m across twelve sub-bands. A four-point dither pattern was used \citep{2015PASP..127..646W,2015PASP..127..584R,2023PASP..135d8003W} as well as dedicated off-source background exposures. The raw, uncalibrated data are reprocessed using the JWST Science Calibration Pipeline version 1.17.1 \citep{2025zndo..17101851B}. During the first stage (\texttt{calwebb\_detector1}), the pipeline performs ramp fitting following corrections for saturation, linearity, dark currents, reset switch charge decay, and cosmic ray jumps. In the second stage (\texttt{spec2}), detector-plane calibrations are executed, which include flat-fielding, straylight correction, flux calibration, background subtraction, and a two-pass fringe correction designed to remove high-frequency static substrate fringes and low-frequency dichroic beating \citep{2023A&A...675A.111A,2023A&A...673A.102G}. In the third stage (\texttt{spec3}), outlier detection and pixel replacement are applied, and the individual dithered frames are combined into twelve rectified \texttt{s3d} data cubes per target using the drizzle algorithm \citep{2023AJ....166...45L}. These resulting cubes feature channel-specific spaxel sizes of 0.13$^{\prime\prime}$, 0.17$^{\prime\prime}$, 0.20$^{\prime\prime}$, and 0.35$^{\prime\prime}$, and spectral samplings of 0.8, 1.3, 2.5, and 6.0~nm for channels 1 to 4, respectively.

\subsection{Post-Processing Pipeline and Spatial Alignment}

All subsequent processing steps are performed with a custom, dedicated Python pipeline. In the initial cube preparation, the science and error extension spaxels are converted from surface brightness units to flux density. Wavelength coordinates are converted to the rest frame using optical redshifts from \cite{2024MNRAS.527..281I} and SDSS DR18, with fine-tuning adjustments made via the strong, isolated HI 6--5 (Pfund-$\alpha$) recombination line. Unreliable detector array edge channels are trimmed, and a $3 \times 3$ spatial smoothing boxcar average is applied to each wavelength slice to compensate for the pixel-to-pixel resampling noise introduced by the drizzling process. To align the twelve sub-bands reconstructed on independent grids and to determine the extraction loci, source centroids are determined in each sub-band separately by fitting a 2D Gaussian. These individual positions are replaced with a single, self-consistent consensus coordinate set calculated as the median of the coordinates from the seven sub-bands with the highest signal-to-noise ratio. Residual low-amplitude detector-level fringing is addressed by applying a 1D residual-fringe correction tool to every spaxel spectrum, where the fit in channels 3 and 4 is strictly limited to the dichroic component to protect genuine spectral structures from being artificially absorbed. To eliminate lingering background pedestals, the median value of a surrounding annulus (placed 2 spaxels beyond the extraction radius and spanning a thickness of 5 spaxels) is subtracted from each wavelength slice, though this step is skipped for highly extended sources to prevent self-subtraction.

\subsection{Spectral Extraction and Characterization of Source Extent}\label{sec:appa3}

Spectral extraction is performed differently depending on the source spatial extent. Source sizes are first characterized without shape assumptions by calculating curve-of-growth half-light radii on bright \Siv\ and \Neiii\ emission-line maps, thereby sorting targets as point-like, compact, or resolved based on a $3\sigma$ threshold comparison with a smoothed point-spread function (PSF) model. 

For point-like and compact sources, we perform for each wavelength slice:
\begin{itemize}
\item A simple aperture extraction whose radius scales with wavelength following the PSF properties. For compact sources, we enlarge this aperture by accounting for the wavelength-independent physical extent.
\item An optimal extraction which scales a 2D model profile to the IFU data at the source centroid location. For point sources the model is the PSF and for compact sources it is the PSF broadened by a narrow Gaussian component whose width is the physical extent.
\end{itemize}
\noindent We use the optimal extraction as our default, since it provides the best S/N (the extraction being weighted by the S/N in each spaxel). We verified that the two methods provide compatible fluxes (i.e., the optimal extraction fluxes should be within the lower S/N aperture extraction error bar).

For extended, irregular, sources whose shape is not well modeled by a broadened PSF, we use an aperture radius ensuring that the 90\% of the line fluxes are enclosed (same radius for all lines). No point-source aperture correction is applied, which makes the absolute fluxes conservative lower limits; however, because the aperture captures a nearly constant fraction of the encircled energy (85--87\%) across all wavelengths, the flux deficit cancels out almost perfectly in emission-line ratios.

All these methods provide us with a 1D spectrum that we cross-calibrate and from which we then measure the lines fluxes. 

\subsection{Cross-Calibration, Line Fitting, and 3D Bayesian Verification}

To resolve spectrophotometric discontinuities of up to 20\% at sub-band boundaries arising from aperture limits, different spatial samplings, and calibration uncertainties \citep{2025AJ....169...67L}, a relative cross-calibration is performed. The pipeline solves for a multiplicative factor per sub-band simultaneously through a global weighted least-squares optimization, constraining the continuum to lie on a single smooth cubic B-spline after masking major emission lines. 

Emission-line fluxes are measured on the calibrated 1D spectra (Sect.\,\ref{sec:appa3}) using non-linear least-squares fitting \citep{2014zndo.....11813N}. The instrumental line-spread function (LSF) is modeled as a concentric double Gaussian calibrated against high-quality reference lines to capture both the core and the shallow, extended wings, and is combined in quadrature with a thermal Doppler width. Fits are computed over a window of $\pm5$ LSF FWHM around the expected line centers. 


\section{Observed line fluxes and ratios} \label{apdx_obslnrts}

In Table \ref{table:calculated_ratios} we present the mid-IR diagnostic emission line ratios used in Figs. \ref{IR-diganostics-master}, \ref{SSP_shock_blending}, and \ref{IMBH-models}. We also provide the observed line fluxes used to derive these ratios in Table \ref{table:calculated_fluxes}. 
In both tables we report 3$\sigma$ limits in case of non-detections.

\begin{table*}[!h]
\caption{JWST/MIRI (MRS) mid-IR emission line flux ratios.}
\label{table:calculated_ratios}
\centering
\small
\begin{tabular*}{\textwidth}{@{\extracolsep{\fill}} l ccccc }
\hline\hline
Object & \Siv/\Siii\ & \Nev/\Neiii\ & \Neiii/\Neii\ & \Oiv/\Siii\ & \Oiv/\Neiii\ \\
\hline
J0811+4730      & 0.55 $\pm$ 0.11 & <-1.28 & >1.16 & <1.26 & <0.70 \\
J1004+3256      & >0.84 & <-1.01 & >1.19 & -- & <0.60 \\
J104458+03531   & 0.69 $\pm$ 0.04 & -2.11 $\pm$ 0.13 & 1.50 $\pm$ 0.05 & 0.05 $\pm$ 0.06 & -0.52 $\pm$ 0.06 \\
J1155+5739      & 0.55 $\pm$ 0.03 & <-2.85 & 1.51 $\pm$ 0.07 & -0.58 $\pm$ 0.06 & -1.08 $\pm$ 0.06 \\
J120202+54155   & 0.49 $\pm$ 0.07 & <-1.43 & 0.91 $\pm$ 0.06 & <-0.22 & <-0.47 \\
J122437+37243   & 0.52 $\pm$ 0.07 & <-1.62 & 1.18 $\pm$ 0.07 & <0.28 & <-0.12 \\
J1234+3901      & >0.21 & <-1.09 & 1.11 $\pm$ 0.11 & -- & -- \\
J1505+3721      & 0.70 $\pm$ 0.12 & -1.23 $\pm$ 0.09 & >1.37 & <1.62 & <1.07 \\
J150934+37314   & 0.78 $\pm$ 0.04 & -2.73 $\pm$ 0.14 & 1.45 $\pm$ 0.02 & -0.29 $\pm$ 0.10 & -0.82 $\pm$ 0.10 \\
J160810+35280   & 1.10 $\pm$ 0.03 & <-2.09 & >1.74 & 0.18 $\pm$ 0.09 & -0.55 $\pm$ 0.09 \\
J2229+2725      & >0.86 & <-1.33 & >1.46 & -- & <0.97 \\
\hline
\end{tabular*}
\tablefoot{Dashes indicate that a calculation could not be performed due to absence of the line. Non-detections which are the result of low S/N ($<3$) are marked as upper limits with "<". In these cases, the fluxes are substituted by their 3$\sigma$ uncertainty limits.}
\end{table*}

\begin{table*}[!h]
\caption{JWST/MIRI (MRS) mid-IR emission line fluxes.}
\label{table:calculated_fluxes}
\centering
\small
\begin{tabular*}{\textwidth}{@{\extracolsep{\fill}} l cccccc }
\hline\hline
Object & \Siv\ 10.5\,$\mu$m & \Neii\ 12.8\,$\mu$m & \Nev\ 14.3\,$\mu$m & \Neiii\ 15.6\,$\mu$m & \Siii\ 18.7\,$\mu$m & \Oiv\ 25.9\,$\mu$m \\
\hline
J0811+4730      & 0.78 $\pm$ 0.03 & <0.05 & <0.04 & 0.79 $\pm$ 0.04 & 0.22 $\pm$ 0.06 & <3.99 \\
J1004+3256      & 1.68 $\pm$ 0.04 & <0.07 & <0.11 & 1.15 $\pm$ 0.04 & <0.24 & <4.58 \\
J104458+03531   & 72.66 $\pm$ 6.59 & 1.73 $\pm$ 0.18 & 0.43 $\pm$ 0.12 & 55.02 $\pm$ 1.30 & 14.99 $\pm$ 0.55 & 16.63 $\pm$ 2.29 \\
J1155+5739      & 295.72 $\pm$ 17.20 & 7.94 $\pm$ 1.20 & <0.37 & 259.98 $\pm$ 9.06 & 83.63 $\pm$ 4.06 & 21.78 $\pm$ 2.71 \\
J120202+54155   & 23.80 $\pm$ 2.84 & 1.69 $\pm$ 0.20 & <0.51 & 13.73 $\pm$ 1.03 & 7.74 $\pm$ 0.73 & <4.62 \\
J122437+37243   & 42.14 $\pm$ 6.02 & 2.12 $\pm$ 0.25 & <0.77 & 32.01 $\pm$ 3.84 & 12.65 $\pm$ 0.92 & <24.35 \\
J1234+3901      & 0.57 $\pm$ 0.09 & 0.05 $\pm$ 0.01 & <0.05 & 0.62 $\pm$ 0.03 & <0.35 & -- \\
J1505+3721      & 2.96 $\pm$ 0.14 & <0.09 & 0.12 $\pm$ 0.02 & 2.06 $\pm$ 0.10 & 0.59 $\pm$ 0.17 & <24.42 \\
J150934+37314   & 159.82 $\pm$ 11.76 & 3.16 $\pm$ 0.13 & 0.17 $\pm$ 0.06 & 89.43 $\pm$ 3.13 & 26.47 $\pm$ 1.19 & 13.54 $\pm$ 3.17 \\
J160810+35280   & 49.77 $\pm$ 1.97 & <0.39 & <0.17 & 21.26 $\pm$ 0.78 & 3.92 $\pm$ 0.18 & 5.97 $\pm$ 1.27 \\
J2229+2725      & 2.03 $\pm$ 0.07 & <0.04 & <0.05 & 1.11 $\pm$ 0.07 & <0.28 & <10.29 \\
\hline
\end{tabular*}
\tablefoot{Dashes indicate that a calculation could not be performed due to absence of the line. Non-detections which are the result of low S/N ($<3$) are marked as upper limits with "<". In these cases, the fluxes are substituted by their 3$\sigma$ uncertainty limits.  All fluxes are in units of $10^{-16} \, \mathrm{erg \, s^{-1} \, cm^{-2}}$.}
\end{table*}

\section{Translation of \Nev/\Neiii\ line ratios between the mid-IR and optical} \label{PyNeb_for_opt_NeV}

\begin{figure}[h]
  \resizebox{\hsize}{!}{\includegraphics{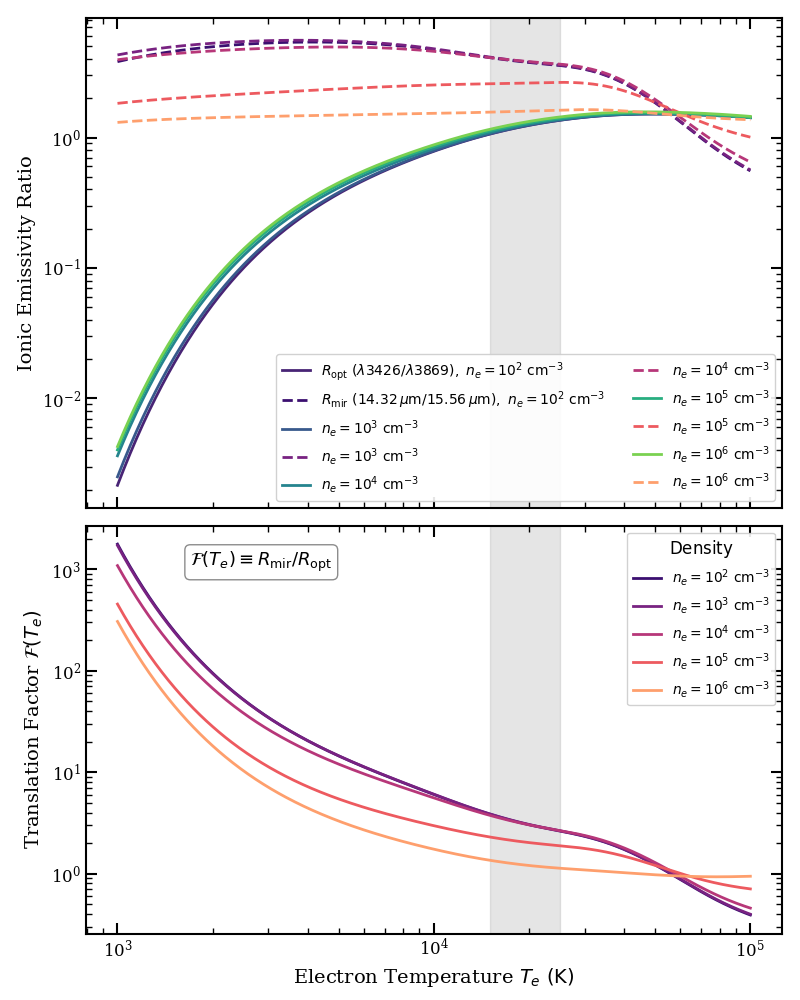}}
  \caption{Theoretical emissivity ratios and the mid-infrared to optical translation factor for $[\mathrm{Ne\,V}]/[\mathrm{Ne\,III}]$ as a function of electron temperature. Top: The emissivity ratio per ion for the optical forbidden transition ($R_{\mathrm{opt}} \equiv j([\mathrm{Ne\,V}]\,\lambda3426) / j([\mathrm{Ne\,III}]\,\lambda3869)$, solid blue curve) compared to the mid-infrared fine-structure transition ($R_{\mathrm{mir}} \equiv j([\mathrm{Ne\,V}]\,14.32\,\mu\mathrm{m}) / j([\mathrm{Ne\,III}]\,15.56\,\mu\mathrm{m})$, dashed red curve). Bottom: The temperature-dependent translation factor, defined as $\mathcal{F}(T_e) \equiv R_{\mathrm{mir}} / R_{\mathrm{opt}}$, required to convert between the optical and mid-infrared diagnostic ratios. All curves were computed using the \texttt{PyNeb} atomic emissivity assuming various densities ($n_e = 10^3-10^6\ \mathrm{cm}^{-3}$). At low electron temperatures ($T_e < 30,000\ \mathrm{K}$), the optical ratio is exponentially suppressed, causing the conversion factor $\mathcal{F}(T_e)$ to scale steeply before approaching unity in the high-temperature coronal regime ($T_e \sim 10^5\ \mathrm{K}$). In both panels, the shaded areas represent the temperature ranges of our EEMPG sample.}
  \label{IR_OPT_Ne_transition_factor}
\end{figure}

To bridge optical spectroscopy and mid-IR observations, we must establish a theoretical calibration between optical and mid-IR neon ionization diagnostics. Using the \texttt{PyNeb} \texttt{Python} package and the CHIANTI atomic database \citep[Version 10.1;][]{1997A&AS..125..149D,2023ApJS..268...52D}, we computed the theoretical ionic emissivity ratios for the optical ($R_{\mathrm{opt}} \equiv j_{\lambda3426}/j_{\lambda3869}$) and mid-IR ($R_{\mathrm{mir}} \equiv j_{14.32\mu\mathrm{m}}/j_{15.56\mu\mathrm{m}}$) lines across an electron temperature grid of $T_e = 10^3 - 10^5\ \mathrm{K}$ at various densities ($n_e = 10^3-10^6\ \mathrm{cm}^{-3}$); 
see Fig.~\ref{IR_OPT_Ne_transition_factor}.
The forbidden optical transitions originate from the excited $^1D_2$ electronic state whose emissivity increases exponentially with $T_e$, whereas the mid-IR fine-structure lines arise from magnetic spin-orbit transitions within the split of ground term and are nearly temperature insensitive. Therefore, their emissivity ratio is exponentially sensitive to the local electron temperature \citep{1985ApJ...291..561D, 2006agna.book.....O, 2013A&A...551A..82S}. This excitation divergence causes the theoretical translation factor $\mathcal{F}(T_e) \equiv R_{\mathrm{mir}} / R_{\mathrm{opt}}$ to decline exponentially with $T_e$ before asymptotically approaching unity in the hot coronal regime. The lower panel presents the conversion factor, $\mathcal{F}(T_e)$, which relates the optical and mid-infrared diagnostics as a function of electron temperature. It shows that over the typical temperature range observed for metal-poor galaxies and our EEMPG sample ($T_e \sim 15000-24000$ K),  $\mathcal{F}(T_e)$ varies by less than a factor $\la 1.6$.


\begin{figure}[h]
  \resizebox{\hsize}{!}{\includegraphics{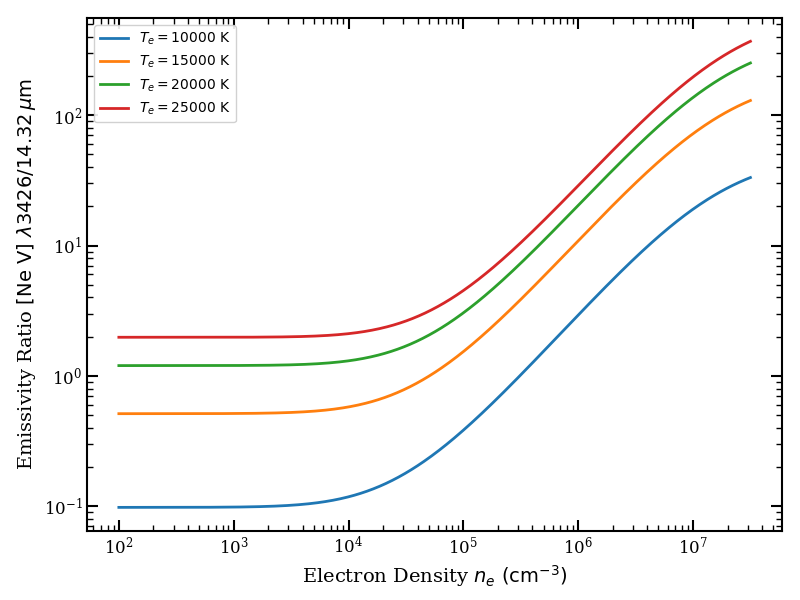}}
  \caption{Emissivity ratio of \Nev$\lambda$3426 \AA\ to \Nev14.32$\mu$m lines as a function of electron density, ($n_e$), for electron temperatures ($T_e$ = 10,000, 15,000, 20,000, and 25,000 K).}
  \label{Ne32_OPT_to_IR}
\end{figure}

In Fig. \ref{Ne32_OPT_to_IR} we see the dependence of the emissivity ratio of the \Nev$\lambda$3426 \AA\ to \Nev14.32$\mu$m lines. For all three temperatures, the emissivity ratio remains relatively flat and constant at electron densities up to approximately 10$^{4}$ cm$^{-3}$. Approaching and beyond the critical density of the \Nev14.32$\mu$m transition, the ratio begins to increase steadily with increasing $n_e$. Additionally, the graph shows that at any given electron density, a higher electron temperature yields a higher emissivity ratio, due to the temperature dependence of \Nev$\lambda$3426.

\onecolumn




\end{appendix}

\end{document}